\documentclass[12pt]{article}

\usepackage{scicite}

\usepackage[style=numeric, sorting=none, firstinits=true, backend=bibtex, defernumbers=true]{biblatex}
\renewbibmacro{in:}{%
  \ifentrytype{article}{}{\printtext{\bibstring{in}\intitlepunct}}}
\usepackage{times}

\usepackage{longtable}
\usepackage{tikz}
\usepackage{pgfplots}
\usepackage{graphics}
\usepackage{caption}
\usepackage{subcaption}
\usepackage{makecell}
\usepackage{threeparttable}
\usepackage{ragged2e}
\usetikzlibrary{arrows, positioning,automata,shapes}
\usetikzlibrary{matrix}
\usepackage{pgf-pie} 
\newcolumntype{H}{>{\setbox0=\hbox\bgroup}c<{\egroup}@{}} 

\newenvironment{sciabstract}{
\begin{quote} \bf}
{\end{quote}}

\newcounter{lastnote}

\title{Innovation with and without patents } 

\author
{Josef Taalbi,$^{1\ast}$ \\
\\
\normalsize{$^{1}$Department of Economic History, Lund University}\\
\normalsize{$^\ast$ E-mail:  josef.taalbi@ekh.lu.se.}
}

\date{}

\begin{document} 

\baselineskip24pt
\maketitle 

\begin{sciabstract}

A long-standing discussion is to what extent patents can be used to monitor trends in innovation activity. This study quantifies the amount and quality of information about actual innovation contained in the patent system, based on 4,460 Swedish innovations (1970-2015) that have been matched to international patents. The results show that most innovations were not patented and that among those that were, 43.9\% of all innovations, only a fraction can be identified with patent quality data. The best-performing models identify 17\% of all information about innovations, equivalent to an information loss of at least 83\%. Econometric tests also show that the fraction of innovations responding to strengthened patent laws during the period were on average 8\% percent. The overlap between the patent and innovation systems is hence more modest than often assumed. This accentuates the need to, alongside patents, develop versatile approaches in order to induce and monitor various aspects of innovation.

\end{sciabstract}

\begin{refsection}
\section*{Introduction}
\label{intro}

Innovation is widely viewed as essential for achieving long-run economic development and sustainability. Measuring and monitoring innovation is therefore key for our understanding of trends in technology and structural transformation over time and between countries. Innovation is typically measured through various indicators covering different stages of the innovation process \cite{dziallas2019}, including scientific output, R \& D expenditures and patents, or compound indices of these \cite{edquist2018, soumitra2020, wipo2021}.

Patents remain the most widely used innovation indicator and for good reason. Patent records are publicly and readily available, contain detailed information about inventors and inventing firms globally and for long time periods. In addition, patent records offer ways of assessing the value of patents through patent citation counts \cite{trajtenberg1990, jaffe2002, hall2005},  and the insights into knowledge flows that can be gained from analysis of patent citations are unparalleled \cite{jaffe2002}. 

There are important sources of discrepancies, however, between innovations and patents, and the extent to which the patent and innovation systems overlap is not settled \cite{fontana2013, higham2021}. The question of how much the patent and innovation systems actually overlap also haunts policy discussions. The patent system has come under fire for being inefficient \cite{stiglitz2007, jaffe2011, derassenfosse2021}, distorting incentives or blocking innovation \cite{heller1998,cohen2016}, while the number of innovations affected by patents or patent laws has been argued to be low or uncertain \cite{moser2005, moser2012, fontana2013, lerner2009}. Gaining a better understanding of this overlap is therefore key also for policy debates.

This study presents evidence about the amount and quality of information on innovation that is possible to identify within the patent system. In doing so, this study addresses methodological issues and supplies evidence for broader discussions about the relationship between the patent and innovation systems. This effort is based on a literature-based innovation output (LBIO) database \cite{taalbi2017what, kander2019} containing 4,460 commercialized innovations in Sweden, one of the world's highest ranked innovative economies \cite{wipo2021}. These innovations, commercialized between 1970 and 2015, were linked to 13,561 patents across various national and supra-national patent offices, through manual and machine-learning-assisted searches in Google Patents. The matching methodology is detailed in the supplementary materials.

\begin{figure*}[h]
\centering
\begin{subfigure}[b]{.50\linewidth}
\begin{tikzpicture}[
 bigcircle/.style={ 
    text width=1.6cm, 
    align=center, 
    line width=1mm, 
    draw, 
    rounded corners, 
     minimum width = 2.5cm, 
    minimum height = 2cm,
    font=\sffamily\footnotesize 
  },
 desc/.style 2 args={ 
  text width=2.5cm, 
  font=\sffamily\scriptsize\RaggedRight, 
  label={[#1,yshift=-1.5ex,font=\sffamily\footnotesize]above:#2} 
  },
 node distance=10mm and 15mm 
]

\node[anchor=north] at (2.2,2.5) {$\rho$ patented};
\node[anchor=north] at (6.2,2.5) {\makecell{recall $\alpha$ \\ with precision $\beta$}};
\node [bigcircle] (circ1) {Innovations};

\node [bigcircle,black,right=of circ1] (circ2) {Patent data};

\node [bigcircle,black,right=of circ2] (circ3) {Identification of patented innovations};
\draw [black!80] (circ1) -- (circ2) -- (circ3) ;
\matrix [
    matrix of nodes, 
draw=black,
line width=0.8mm,
below = of circ1,
    nodes={font=\ttfamily},
    every node/.style={anchor=base,text depth=.5ex,text height=2ex,text width=4em}
]
  {
Skype    \\
Spotify   \\
NMT \\
SBR \\
  };

\matrix [
    matrix of nodes, 
draw=black,
line width=0.8mm,
below = of circ2,
    nodes={font=\ttfamily},
    every node/.style={anchor=base,text depth=.5ex,text height=2ex,text width=4em}
]
  {
    Skype    \\
Spotify   \\
|[fill=black!10]| \\
|[fill=black!10]|  \\
Patent 1\\
Patent 2\\
Patent 3\\
  };

\matrix [
    matrix of nodes, 
draw=black,
line width=0.8mm,
below = of circ3,
    nodes={font=\ttfamily},
    every node/.style={anchor=base,text depth=.5ex,text height=2ex,text width=4em}
]
  {
    Skype    \\
|[fill=black!10]| \\
|[fill=black!10]| \\
|[fill=black!10]|  \\
Patent 1\\
Patent 2\\
|[fill=black!10]| \\
  };
    
    \node[anchor=north] at (2.1,-4) {$\rho = 1/2$ };
\node[anchor=north] at (6.2,-4) {\makecell{$\alpha=1/2$, \\ $\beta=1/3$}};

\end{tikzpicture} 
\end{subfigure}

\caption{(a) Innovations and patents viewed as information transmission through a noisy channel. A fraction $\rho$ of $N$ innovations enter into the patent system. The patent system also contains noise in the form of non-commercialized inventions that can be reduced through patent quality measures. The quality of the information about innovations depends on the fraction of true positives identified (recall $\alpha$), and the probability that a patent identified is truly an innovation (precision $\beta$). (b) Example. If asked to name the patented innovation, the patent analyst would on average be able to correctly identify 1 correctly a third of the times. The information about the original source, as a fraction of the total, is $\rho \times \alpha \times  \beta$ or $1/12$ in the example.} 
\label{fig:information}
\end{figure*}
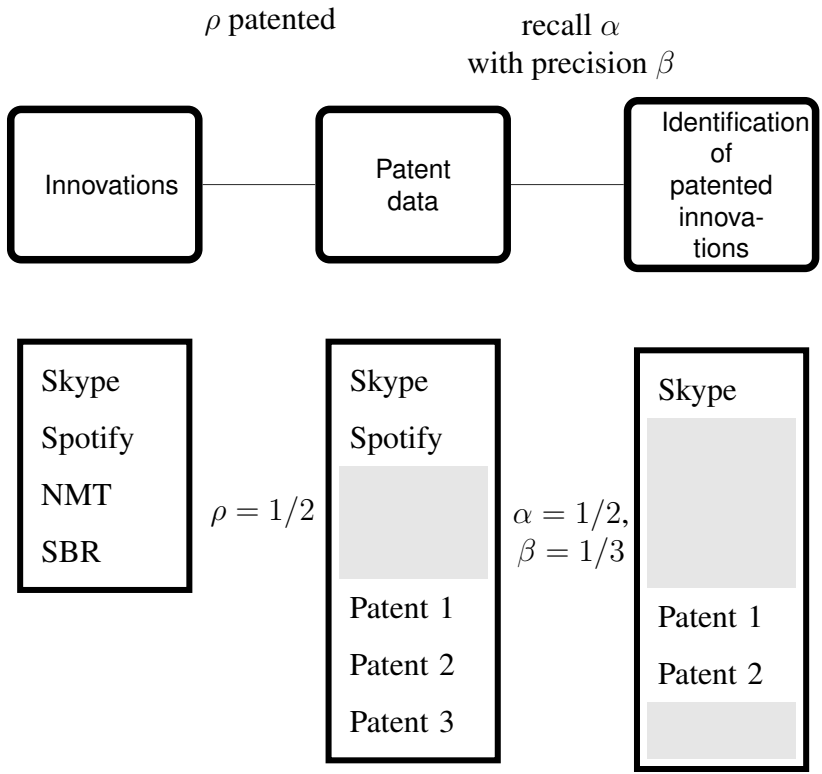

This data enables examination of the information overlap between patents and commercialized innovations from multiple angles. This study views the problem of measuring innovation as analogous to information transmission from a source to a receiver through a noisy channel (Fig. \ref{fig:information}). The key question is how much information the receiver (patent analyst) has about the source (innovations). This study phrases this question as follows: what fraction of innovations can be correctly identified by a patent analyst, based on data available within the patent system?

This fraction is circumscribed by three factors (Fig. 1). First, not all innovations are patented, but some fraction $\rho$ that is determined by property laws and appropriability strategies. Previous research proposes varying estimates of the percentage of innovations that are patented, from 9.6\% \cite{fontana2013}, 36\% \cite{arundel1998} to almost half \cite{cohen2000}. A second important issue is that patents, strictly speaking, reflect invention, which may, or may not, lead to \emph{innovation}, viz. new combinations that are commercialized or otherwise come into economic use \cite{oecd2018}. Patenting is also often the outcome of strategic decisions to protect intellectual advances, rather than reflecting innovation activity \cite{dernis2001}. Therefore, to weed out less important or less valuable patents, the usual strategy is to use patent citations. However, while some studies suggest patent citations to be a good measure of economic value \cite{trajtenberg1990, jaffe2002,hall2005}, others find that patent citation counts are ``noisy'', heterogeneous over time, across sectors and countries \cite{criscuolo2008, gambardella2008, roach2013, higham2021, lerner2022}. Such quality-adjusted indicators must balance two aspects: the amount of actual innovations (true positives) that are captured, the recall $\alpha$, and the fraction of true positives among all patents identified,  the precision $\beta$. The information about innovations in the final patent selection is then defined by the fraction of innovations covered and the precision of the selection:  $\rho \times \alpha \times \beta $. This measure is further motivated in the supplementary materials.

\section*{Results}

\paragraph*{Patent propensity}

\begin{figure*}[h]
\begin{subfigure}[b]{.45\linewidth}
\caption{}
\label{fig:patprop_cum}
\includegraphics[width=7cm, height=7cm, keepaspectratio]{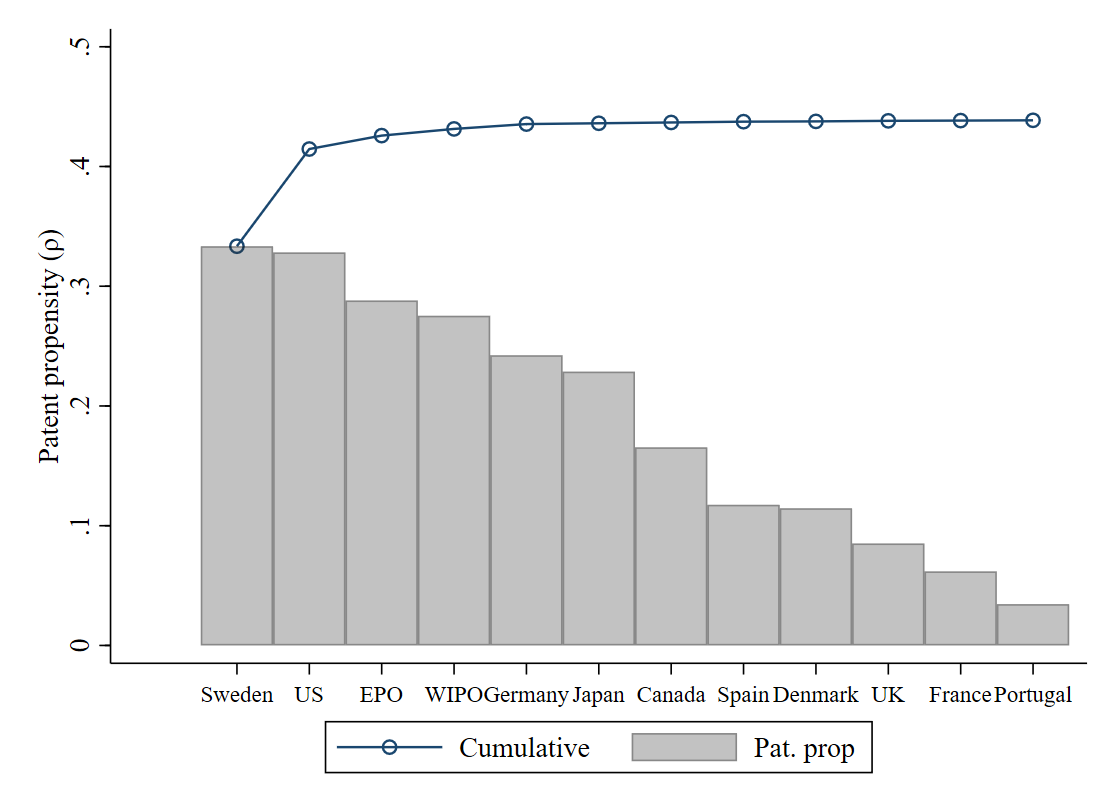}
\end{subfigure}
\begin{subfigure}[b]{.50\linewidth}
\centering
\caption{}
\label{fig:proptot}
\includegraphics[width=7cm, height=7cm, keepaspectratio]{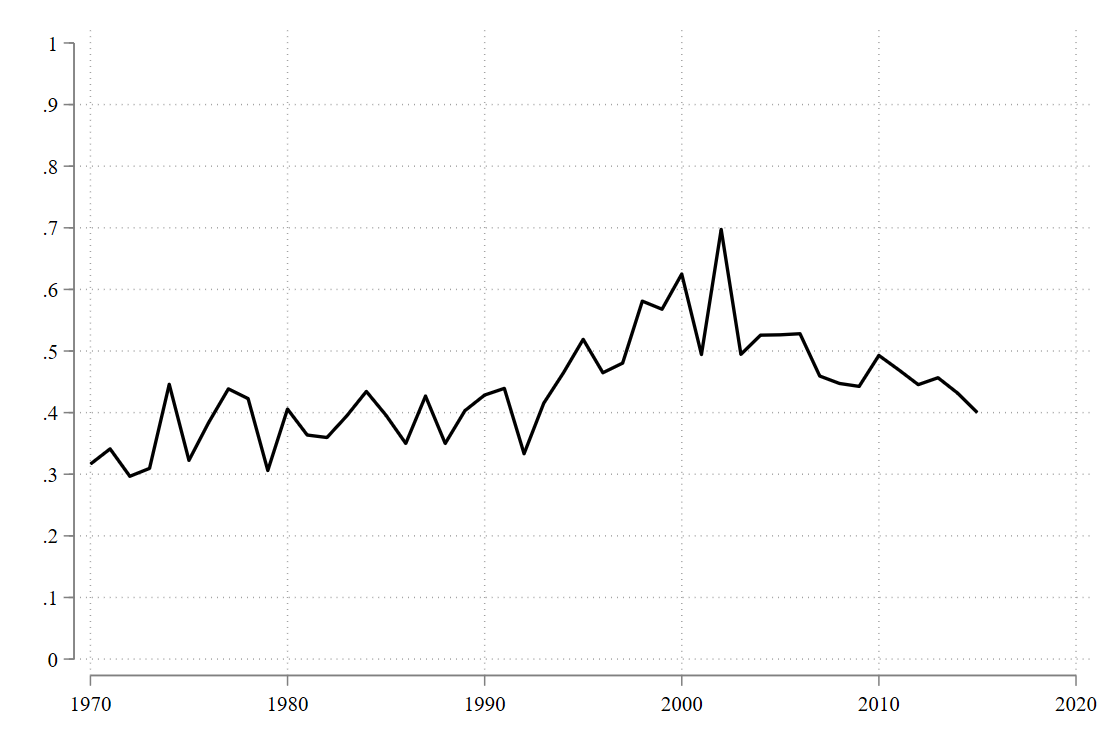}
\end{subfigure}
\begin{subfigure}[b]{.50\linewidth}
\caption{}
\label{fig:propPO}
\includegraphics[width=7cm, height=7cm, keepaspectratio]{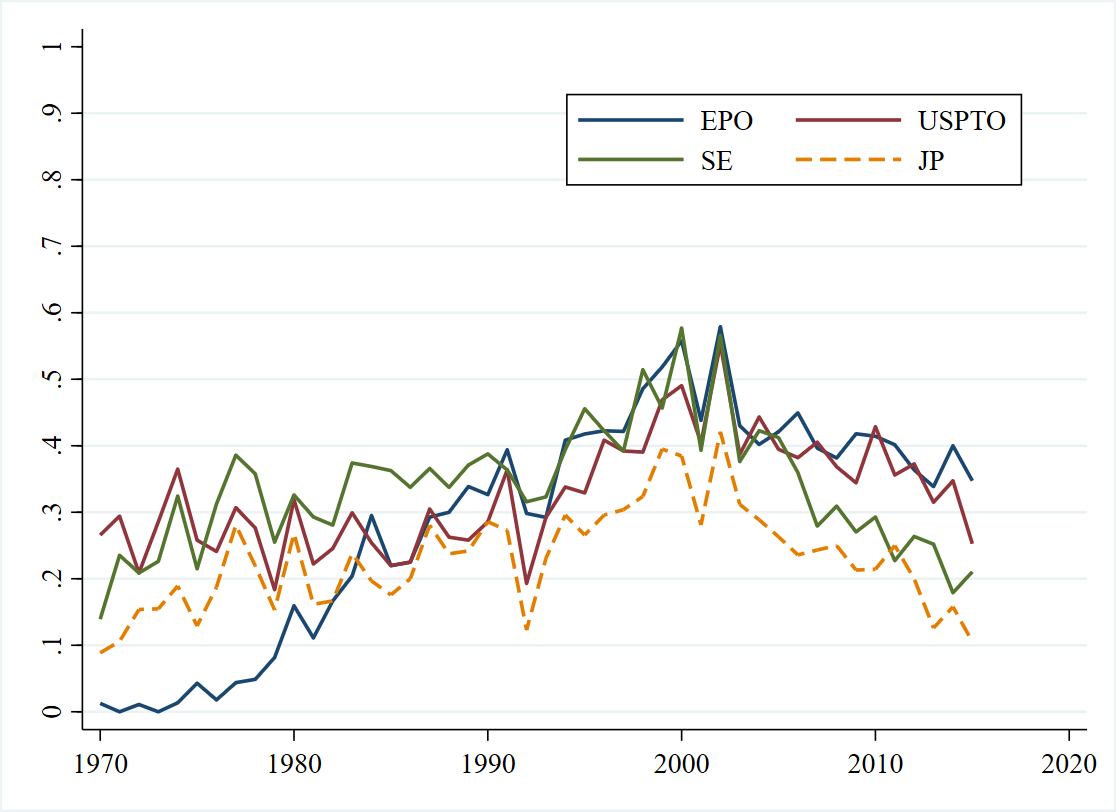}
\centering
\end{subfigure}
\begin{subfigure}[b]{.50\linewidth}
\centering
\caption{}
\label{fig:sectoral}
\includegraphics[width=7cm, height=7cm, keepaspectratio]{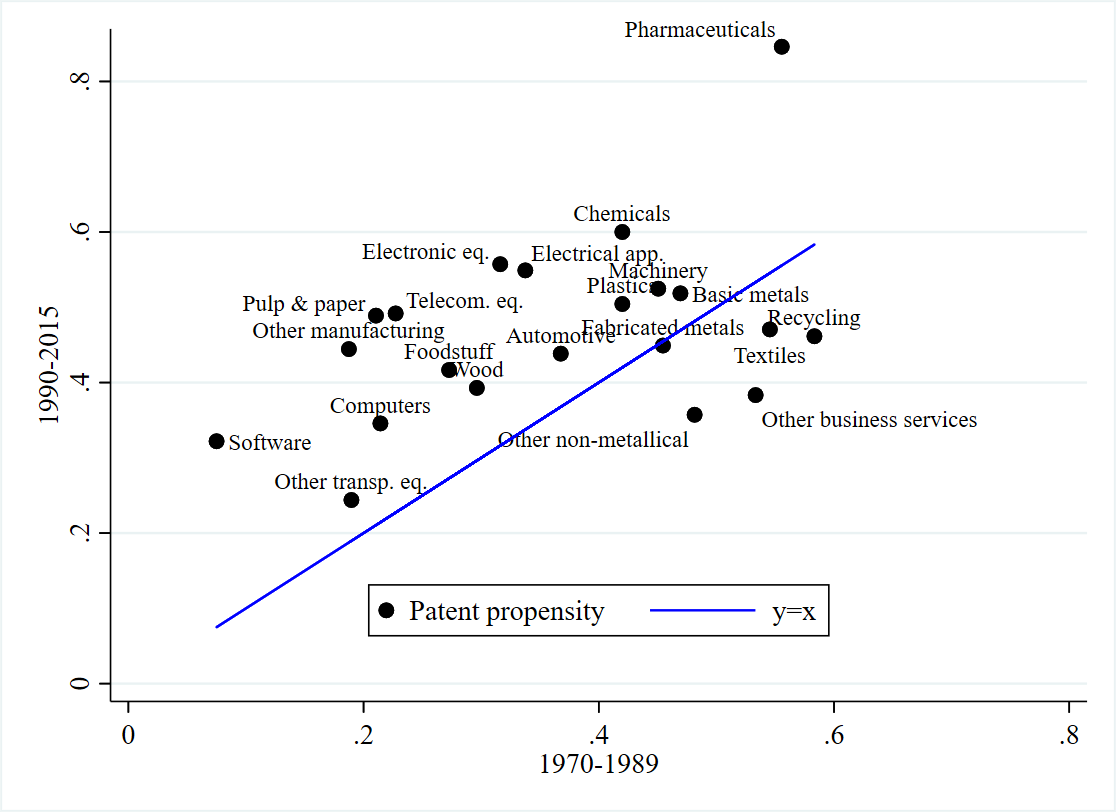}
\end{subfigure}
\caption{(a) Patent propensity by patent office. (b) Total patent propensity, by commercialization year 1970-2015, (c) EPO, USPTO, Sweden, Japan, (d) Patent propensity across sectors (ISIC Rev. 3), all patent offices. Note: Results not given when sectoral counts are below five.}
\label{fig:breakdown}

\end{figure*}

\begin{figure*}[h]
\begin{subfigure}[b]{.50\linewidth}
\centering
\caption{}
\label{fig:decomposition}
\includegraphics[width=7cm, height=7cm, keepaspectratio]{decomposition.png}
\end{subfigure}
\begin{subfigure}[b]{.50\linewidth}
\centering
\caption{}
\label{fig:counterfactual}
\includegraphics[width=7cm, height=7cm, keepaspectratio]{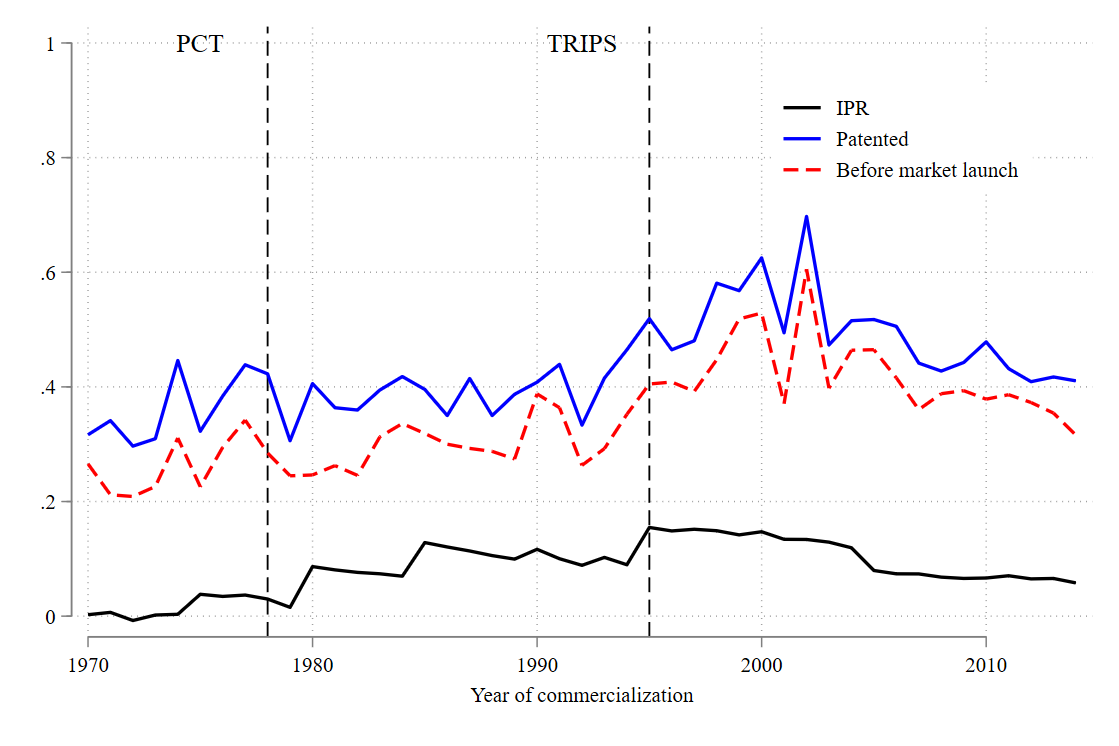}
\end{subfigure}
\caption{(a) Decomposition of changes in average patent propensity from previous decade. Percentage point contributions from changes within sectors, between sectors and an interaction effect. (b) Patent propensity and estimates of upper bound of the share of innovations dependent on IPR policy changes for five countries.}
\label{fig:patentpropensity}
\end{figure*}

We turn first to the propensity to patent ($\rho$ in Fig. \ref{fig:information}). The main results are given in Fig. \ref{fig:breakdown} and further detailed in Table \ref{tab:sumstat}. The results show that 43.9 percent of all innovations, launched in 1970-2015, were patented in at least one patent office, whereas patenting propensity to any one single patent office was highest for the Swedish and US patent offices. Combining data from two or three patent offices however suffices to capture a near-complete set of all patented innovations (Fig. \ref{fig:patprop_cum}). 

Piercing below the aggregate reveals stark differences over time and across sectors. The patent propensity to the Swedish Patent Office has increased from a low level of 13.9\% in 1970 to 57.7\% in 2000, followed by a decrease back to 21.1\% in 2015. The patent propensity to USPTO increased from 26.6\% in 1970 to 49.0\% in 2000, followed by a decrease back to 25.3\% in 2015 (Fig. \ref{fig:propPO}). These results are in line with the emergence of a pro-patent era in the late 1980s \cite{granstrand2000, granstrand2012}.

Some studies have suggested that this may have been driven by an increased emphasis on high-tech industries where patents are an important means of rent appropriation \cite{kim2004}. Table \ref{tab:sectoral} and Figure \ref{fig:sectoral} suggest that patenting propensity differs substantially across sectors. In line with earlier studies \cite{arundel1998}, we see that high-tech industries like R\&D services and pharmaceuticals have had an especially high patent propensity. 
At the other side of the spectrum, several industries, including paper and pulp, foodstuff, wood, and ICT sectors like computer equipment and software, all have had a patenting propensity below 40\% and some below 30\%. These differences are statistically robust in logistic regressions that predict the propensity to patent an innovation (Table \ref{tab:logistic}). At the level of individual innovations, we also observe generic statistical associations between the developmental complexity of an innovation, its radicalness and the propensity to patent it (Table \ref{tab:logistic}). 

Meanwhile, it is also evident from Fig. \ref{fig:sectoral} that patent propensity has increased in most sectors, including low-tech sectors. To better understand the drivers of these patterns, a simple decomposition was carried out of the change of average overall patent propensity between decades (1970s, 1980s, 1990s, 2000s and 2010s). It is clear from this analysis (Figure \ref{fig:decomposition}) that the trends observed mainly reflect generic changes in the propensity to patent, and that the patterns are not driven by any one especially patent intensive sector. In fact, during the decades when patent propensity increased, the ``between effect'' was negative, suggesting that growing sectors tended to be sectors with relatively low patent propensity. This is in contrast to the notion that the patent intensive ICT sectors drove increases in patent propensity \cite{kim2004, holgersson2018}.

As these patterns are generic, one must look for structural explanations. For this reason, econometric tests were carried out, analyzing whether the propensity to patent a Swedish innovation in a given country depends on the country's patent laws, including patent duration, coverage, and enforcement \cite{ginarte1997,park2008}. Applying a multivariate probit approach, the findings (Fig. \ref{fig:counterfactual} and Table \ref{tab:mvprobit}) confirm that these patterns are partially explained by strengthened international patent laws, affecting all industries. According to these results, 8\% of all innovations can be linked to strengthened patent laws, since 1970 (Fig., \ref{fig:counterfactual}). After the TRIPS agreement, the estimated effect was 9.8\%, reaching at most a yearly percentage of 15.5\% in the late 1990s.

These estimates can also be understood as an upper bound of the fraction of innovations that were forthcoming due to strengthened intellectual property rights (see supplementary text for further discussion). Interpreted as such, the results indicate a limited impact of patent laws on innovation in line with previous work \cite{moser2005, moser2012, lerner2009}.

\paragraph*{Prediction of innovations from patent quality statistics}

\begin{figure*}[htbp]
\centering
\footnotesize
\begin{subfigure}[b]{.45\linewidth}
\caption{}
\label{fig:precisionrecall}
\includegraphics[width=7cm, height=7cm, keepaspectratio]{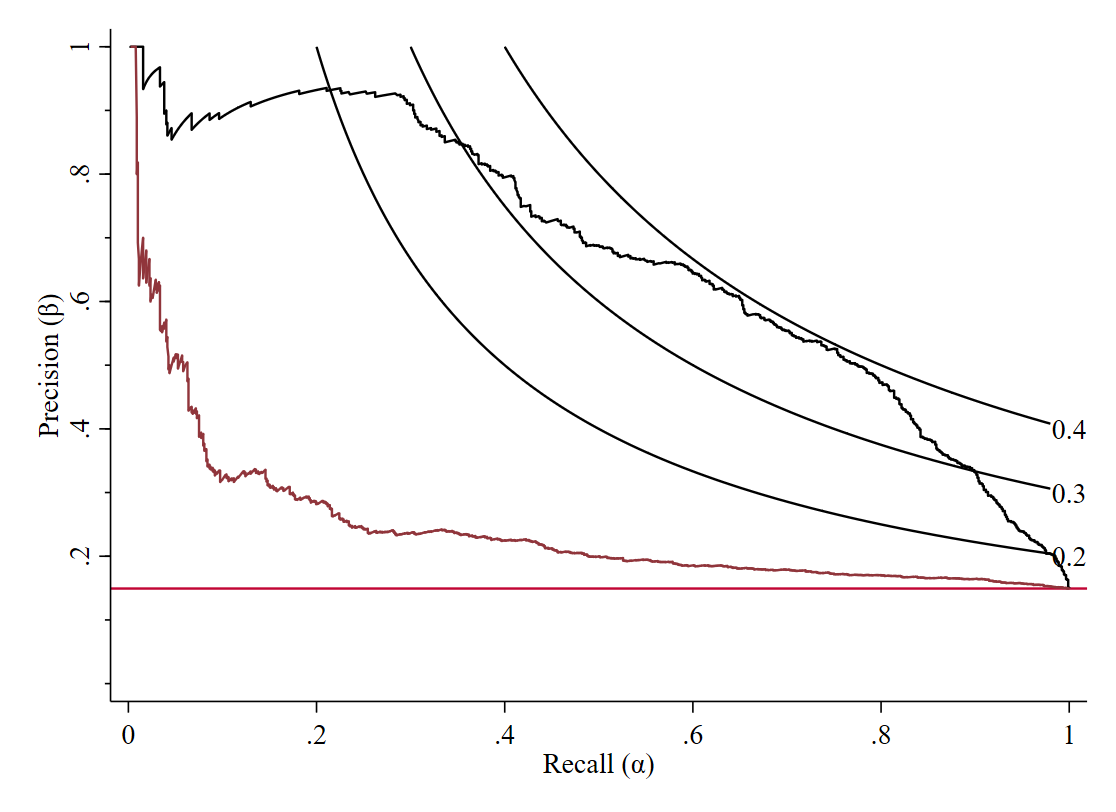}
\end{subfigure}
\begin{subfigure}[b]{.45\linewidth}
\centering
\footnotesize
\caption{}
\label{fig:stripplot_cits}
\includegraphics[width=7cm, height=7cm, keepaspectratio]{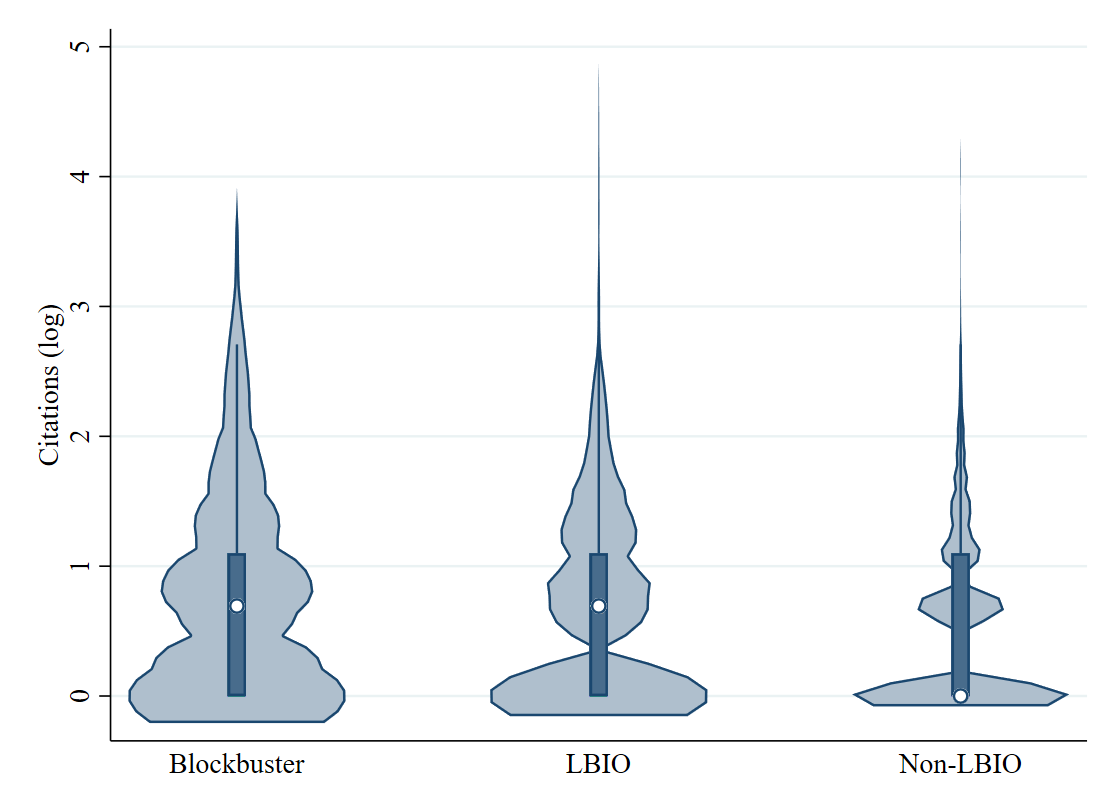}
\end{subfigure}
\begin{subfigure}[b]{.45\linewidth}
\centering
\caption{}
\label{fig:sshare_cits}
\footnotesize
\includegraphics[width=7cm, height=7cm, keepaspectratio]{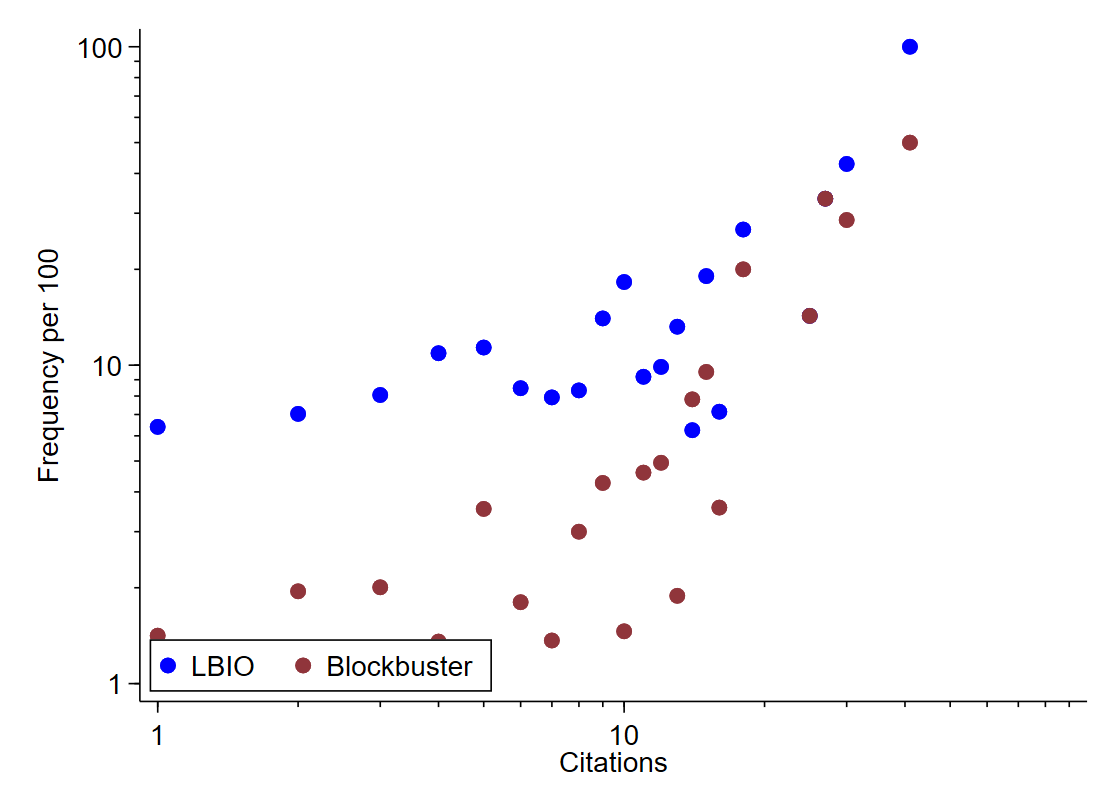}
\end{subfigure}
\begin{subfigure}[b]{.45\linewidth}
\centering
\footnotesize
\caption{}
\label{fig:frequency_cits}
\includegraphics[width=7cm, height=7cm, keepaspectratio]{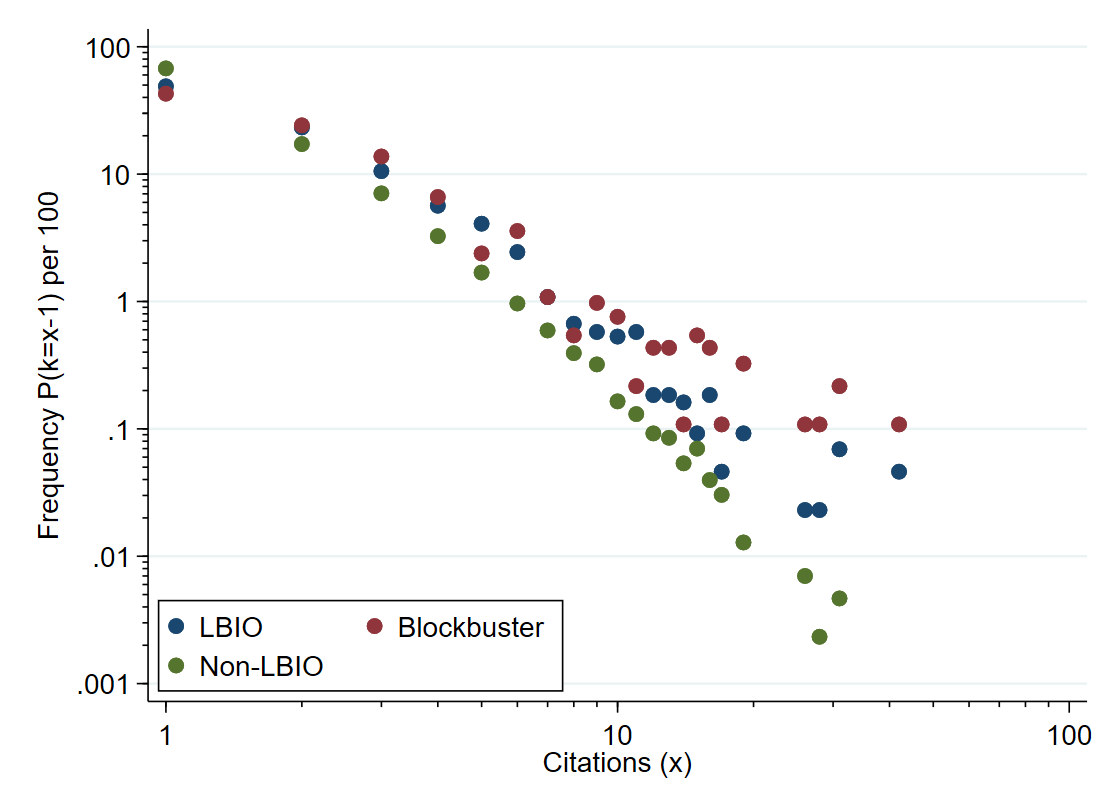}
\end{subfigure}
\begin{subfigure}[b]{.45\linewidth}
\centering
\footnotesize
\caption{}
\label{fig:PCscatter}
\includegraphics[width=7cm, height=7cm, keepaspectratio]{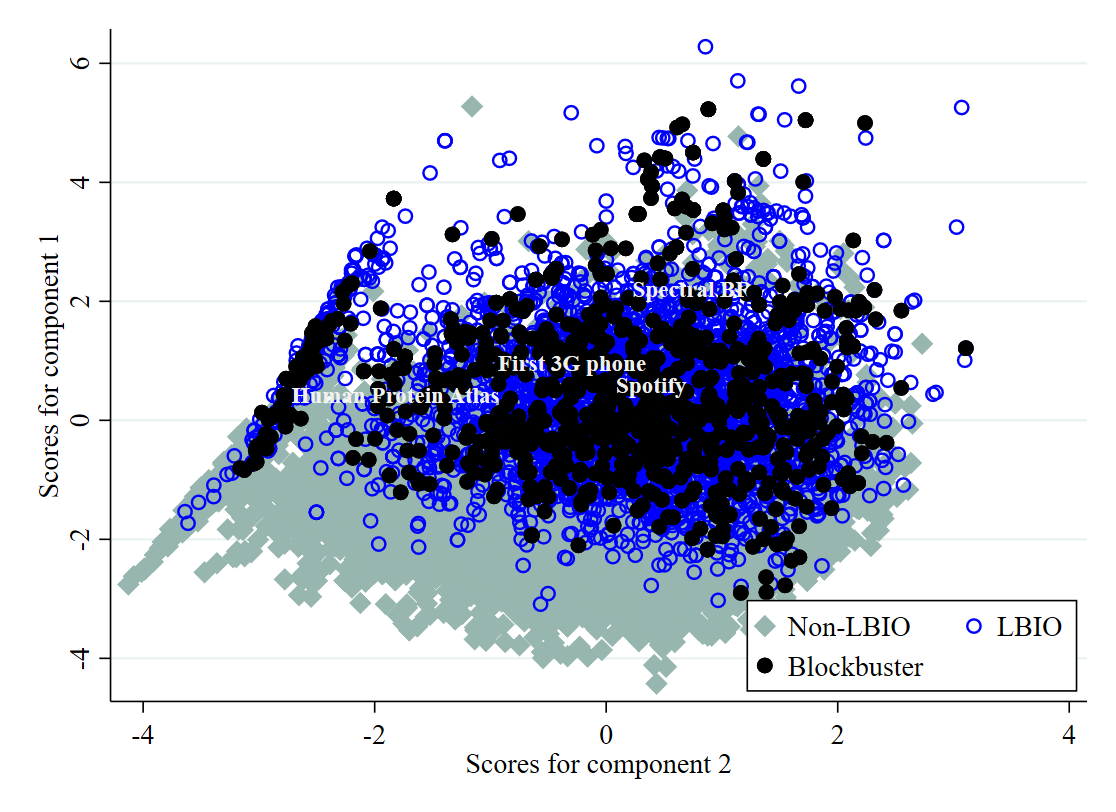}
\end{subfigure}
\begin{subfigure}[b]{.45\linewidth}
\centering
\footnotesize
\caption{}
\label{fig:stripplot_pc1}
\includegraphics[width=7cm, height=7cm, keepaspectratio]{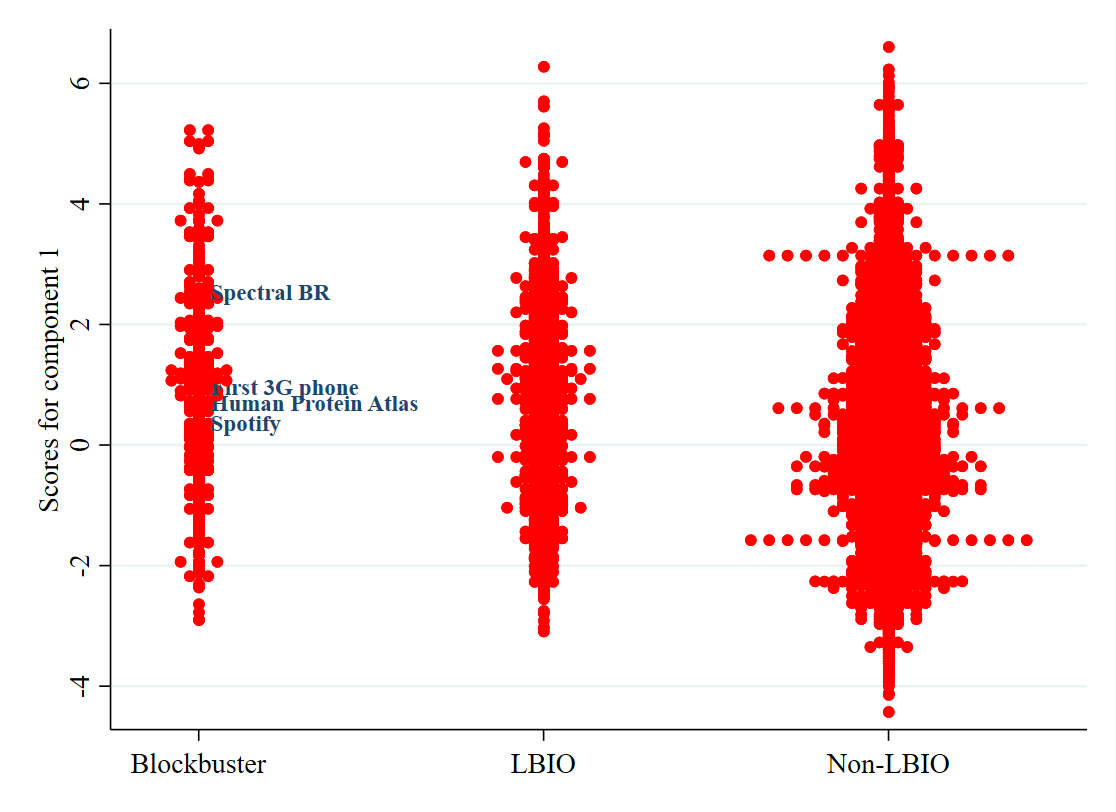}
\end{subfigure}

\caption{(a) Precision-recall for predicting blockbuster innovations, with and without controls for technology fields. 
(b) Violin plot of patent citations within 7 years (log scale) with minimum, first quartile, median, third quartile and maximum, (c) share of patents that have link to LBIO or blockbuster innovations, by number of citations. (d) Frequency distribution of patent citations within 7 years, (e) Principal components for Non-LBIO, LBIO and blockbuster groups of patents, (f) Distribution of first principal component for the Non-LBIO, LBIO and blockbuster subsets.}

\end{figure*}

The second point of interest is to what extent patent citations and other patent quality measures can, under ideal circumstances, be used to separate out radical innovations from less radical innovations. To this end, this study analyzes how well patent quality measures predict four different measures of significant innovation (supplementary materials, Fig. \ref{fig:precisionrecall_all} and Tables \ref{tab:predictsuper_without}-\ref{tab:predictsuper}). The benchmark data is a set of innovations, here called ``blockbuster innovations'' that compares patents linked to 40 innovations known to have been highly successful, to a set of innovations that are known to have been incremental and of little or no economic importance (the methodology is further described in supplementary materials). This approach assumes that the patent analyst has access to training data or has knowledge of the parameters for patent quality data from the OECD Patent Quality Indicators database \cite{squicciarini2013} and heterogeneity across technology fields. The patent quality variables are the number of citations received in the 7 years since the patent was published, patent renewal, family size originality and radicalness \cite{squicciarini2013}, further detailed in supplementary methods and data. Using EPO patent data linked to these innovations, logistic regressions are estimated to find the best performing models. 

The findings suggest that while there is a robust positive correlation between patent quality measures and significant innovations, this association is noisy and has low predictive power without the addition of controls for technology fields. In the best case uncovered with the present data, the models could identify a fraction $\alpha =0.525$ of all patented innovations, with a precision of $\beta=0.753$ (Fig. \ref{fig:precisionrecall}). Predictions for other definitions of significant innovations have similar or poorer performance (Fig. \ref{fig:precisionrecall_all} and Table \ref{tab:predictsuper_without}-\ref{tab:predictsuper}). 

Descriptive statistics are informative as to the somewhat weak discriminatory power of patent quality measures. Descriptive statistics (Table \ref{tab:citations}) show that patents not linked to the innovation database have a mean number of citations of 0.761. Innovations have on average more than double the number of citations, while the blockbuster innovations have, on average, only a marginally higher number of citations than the average LBIO innovation. Fig. \ref{fig:frequency_cits} shows that LBIO and blockbuster innovations tend to have higher patent citations. Patents connected to the LBIO database are skewed to the right, as are the blockbuster innovations. Moreover, the higher number of patent citations a patent has, the more likely a patent is to be linked to an innovation (Fig. \ref{fig:sshare_cits}).  Similar patterns hold for the other patent quality measures. To illustrate, Fig. \ref{fig:PCscatter} shows principal components of the patent quality measures for LBIO, blockbuster innovations and non-LBIO patents. Fig. \ref{fig:stripplot_pc1} shows the distribution for the first principal component. These results show that there is considerable overlap in the distribution of the three sets, but it is also clear that LBIO and blockbuster innovations score higher on average. 

Meanwhile, Fig. \ref{fig:frequency_cits} also shows that the vast majority of patents, including patents linked to blockbuster innovations, have a fairly low number of citations. 72\% of all innovations and 67\% of all blockbuster patents had zero or only one citation. Considering the above, one must conclude that patent citation counts is a noisy indicator of significance \cite{criscuolo2008}. In other words, the results suggest that patents with high citation counts tend to capture significant innovations. However, the converse does not hold: a low patent citation count does not imply that a patent was insignificant (compare \cite{abrams2018}).

\paragraph*{Information loss}

Taken together, an optimistic estimate of the information content about actual innovations that can be identified from within the EPO patent system is a fraction 0.13 of the total information in the case of innovations commercialized in 1977-2015. Using two or three patent office sources could increase the expected information content to 0.17. These figures however, make the assumption that correct controls can be used, accounting for variations in patenting behavior across technology classes and over time. This is implausible without some type of training data. The results support the notion that measuring innovation requires going beyond single dimensions of innovation indicators.

\section*{Discussion}

The present study offers new evidence about the overlap in the patent and innovation systems for one of the world's most innovative economies, 1970-2015. A general result is that patents must be viewed as offering a glance of innovation activity, but the overlap between the patent and innovation systems is limited. Some studies \cite{fontana2013} have warned about the risks of using patent data without a sound grasp of the extent to which patent provide a partial and perhaps biased representation of innovation activity. The results of this study are in certain respects lenient, especially concerning patent propensity in the most recent decades, in sectors where patents are the dominant means of appropriation, and especially if one considers the overall patenting level. The importance of patenting has increased over time with increased international patent law stringency, and the results as regards patent propensity are close to some of the more optimistic figures reported previously \cite{arundel1998, cohen2000}.  

On the other hand, the results suggest that available patent quality indicators on their own are unlikely to be robust proxies of the quality of innovations.  With the use of machine learning \cite{lerner2022} and training data, such as commercialized innovations and appropriate controls for sectoral and temporal variations in patenting behavior, models can be fitted to achieve better results. However, if the goal of the analysis is to capture more significant innovation activity, even the here idealized circumstances suggest an information loss of at least 83\%. 

Overall, these results have implications for our use of patent data, and clearly indicate the necessity of a versatile approach to innovation. Available innovation indicators, cover several steps in the innovation process, from scientific production, R \& D expenditure and patenting activity. In this regard, one must emphasize that patent data remains unparalleled as a window on inventor activity, firm collaborations and non-commercialized invention processes. Similarly, patent citation data remains unparalleled as a window on knowledge flows. While there is great variety in indicators in the pipeline of innovation processes \cite{dziallas2019}, a missing link is consistent, long-term, data on innovation output. Self-reported innovation data, such as the community innovation surveys \cite{arundel2013}, offer one important alternative for recent years. The innovation database used in this study is based on trade journal sources and is one of two long-term innovation output databases to date \cite{kander2019,have2009}. Policy makers and researchers interested in monitoring long-term innovation trends should investigate ways of constructing consistent innovation indicators. One possible cost efficient route is to use alternative methodologies, such as literature-based innovation with patent matching, to train matching algorithms in patents. The results of this study suggest, however, that these exercises still risk being imprecise. Another route, though more costly and time-consuming, implies prioritizing the construction of similar long-term databases based on expert opinion, trade journals, or other literature.

There are also clear implications for innovation policy. The fraction of innovations responding to strengthened patent laws during the period were on average 8\% percent, peaking at 15.5\% in the 1990s. Although there is no consensus, patent policy instruments may have beneficial effects on knowledge accumulation, but should, if the current results are generalizable, not be viewed as a substitute for innovation policy.

\printbibliography

\section*{Acknowledgments}
 I wish to thank Markus Isaksson, Mathias Johansson, Asli Kilicaslan, and Jakob Nyqvist for excellent research assistance, and Frank van der Most for creating the infrastructure that made the patent matching process possible. I also thank Walter G. Park for generously sharing data on intellectual property rights.
 \textbf{Funding:} I gratefully acknowledge funding support from the Sweden's governmental agency for innovation systems, Vinnova (grant no 2020-01963).

\end{refsection}
 \renewcommand{\theequation}{S\arabic{equation}}
\setcounter{equation}{0}
\setcounter{table}{0}
\setcounter{figure}{0}
\renewcommand{\thefigure}{S\arabic{figure}}
\renewcommand{\thetable}{S\arabic{table}}

 \clearpage
\begin{refsection}
\section*{Supplementary materials}
Materials and Methods\\
Supplementary Text\\
Figs. \ref{fig:matchingprocess} to \ref{fig:precisionrecall_all} \\
Tables \ref{tab:sumstat} to \ref{tab:rounds-stats}\\
References \textit{(38-72)}

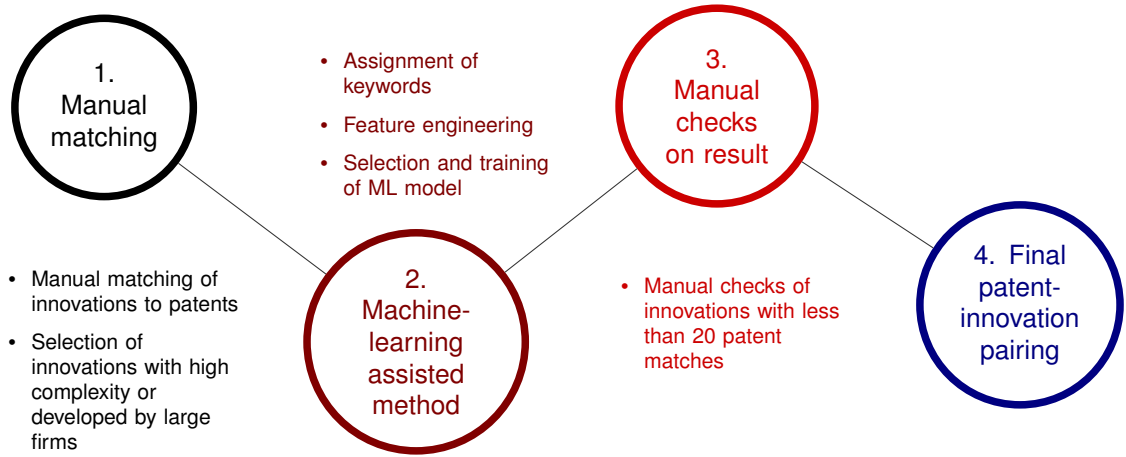
\begin{figure*}[hptb]
\centering
\begin{tikzpicture}[
 bigcircle/.style={ 
    text width=1.6cm, 
    align=center, 
    line width=1mm, 
    draw, 
    circle, 
    font=\sffamily\footnotesize 
  },
 desc/.style 2 args={ 
  text width=4cm, 
  font=\sffamily\scriptsize\RaggedRight, 
  label={[#1,yshift=-1.5ex,font=\sffamily\footnotesize]above:#2} 
  },
 node distance=3mm and 5mm 
]

\node [bigcircle] (circ1) {1. Manual matching};
\node [desc={black}{ },below=of circ1] (list1) {
\begin{itemize}
\setlength\itemsep{0pt} % reduce space between items in list
\item Manual matching of innovations to patents
\item Selection of innovations with high complexity or developed by large firms
\end{itemize}
};

\node [bigcircle,black!50!red,right=of list1] (circ2) {2. Machine-learning assisted method};
\node [desc={black!50!red}{ },text=black!50!red, above=of circ2] (list2) {
\begin{itemize}
\setlength\itemsep{0pt}
\item Assignment of keywords
\item Feature engineering
\item Selection and training of ML model

\end{itemize}
};

\node [bigcircle,black!20!red,right=of list2] (circ3) {3. Manual checks on result};
\node [desc={black!20!red}{},text=black!20!red, below=of circ3] (list3) {
\begin{itemize}
\setlength\itemsep{0pt}
\item Manual checks of innovations with less than 20 patent matches
\end{itemize}
};

\node [bigcircle,black!50!blue,right=of list3] (circ4) {4. Final patent-innovation pairing};
\node [desc={black!50!blue}{},text=black!50!blue, above=of circ4] (h) {

};

% draw the line between circles
\draw [black!80] (circ1) -- (circ2) -- (circ3) -- (circ4);
\end{tikzpicture} 
\caption{Overview of matching process}
\label{fig:matchingprocess}

\end{figure*}

\subsection*{Materials and methods\footnote{The matching methodology is further described in an unpublished working paper \cite{johansson2022}. Figs. \ref{fig:matchingprocess} and \ref{fig:MLmatching}-\ref{fig:zipf} and Tables \ref{tab:features}-\ref{tab:rounds-stats} reproduced here with consent.}}
\label{sec:methods}

\paragraph*{Innovation data}
A contribution of this paper lies in combining a Swedish innovation output database (SWINNO) with patent data for the period 1970-2015 \cite{Sjoo2013SI,kander2019SI,taalbi2021SI}. The data describes a nationwide longitudinal matching between literature-based innovation output and patents \cite{arundel1998SI, brouwer1999SI, cohen2000SI, fontana2013SI, kleinknecht2002SI}. 

The innovation output database is based on the literature-based innovation output (LBIO) methodology, proposed in the late 1980s, owing in part to the recognition of methodological issues with patents and R\&D as innovation indicators \cite{Kleinknecht1993}. The major advantages of the LBIO approach include capturing commercialized innovations, as well as enabling long-run time series of innovation activity in firms, industries or countries. There is now a sizeable number of studies based on this approach, including studies on firms or industries in Spain, Japan, Netherlands and the US \cite{acs1990SI, alegrevidal2004SI, coombs1996SI, edwards1984SI, greve2003aSI, Grawe2009SI, santarelli1996SI, Sjoo2013SI, walker2002SI}.  
However, to date, national databases with a long-run ambition and comprehensive industry coverage exist only for Sweden \cite{Sjoo2013SI, kander2019SI, taalbi2017whatSI, taalbi2021SI} and Finland \cite{palmberg1999SI, Saarinen2005SI, makkonen2013SI, kander2019SI}.

The LBIO database is based on a selection of 15 trade journals covering product innovations and processes sold on a market for the manufacturing industry and ICT services. The database currently covers in total about 1,200 innovations launched between 1908 and 1969 \cite{taalbi2021SI} and about 4,800 innovations from 1970 to 2019 \cite{kander2019SI}. The trade journals are independent from private interests and screened for articles edited by journalists who have made an editorial choice to include an innovation because of its novelty or significance. Hence, the database captures commercialized innovations, whose degree of novelty or significance was clearly stated in independent trade journal articles.

\paragraph*{Matching methodology}

The matching and linking of a patent to an innovation was made by the SWINNO research team \cite{johansson2022SI} using Google Patents, a search engine that indexes patents and patent applications from EPO, USPTO, the Swedish Patent Office and other countries’ patent offices. 

Fig. \ref{fig:matchingprocess} gives an overview of the matching process. In a first step, innovations were matched. Innovations developed by small and medium-sized firms and innovations with low complexity were easy to screen and match, and this process was therefore done manually. This is particularly the case for firms that only have filed a small number of patents. The vast majority of LBIO innovations, roughly 3,600, were manually screened. 

\begin{figure*}
\centering
\footnotesize
\begin{tikzpicture}[xscale = 2.5, 
 font=\sffamily,
mystyle/.style={draw=white, row sep=-\pgflinewidth, thick, text=black, font=\sffamily\bfseries},
]
\pie[square,  
style={mystyle},
color={
yellow!20!red, %Manually, No
blue!70!white, %Manually, Yes
yellow!50!red, %ML, No
blue!30!white, %ML+manually, Yes
yellow!80!red,  %ML+manually, no
blue!10!white%ML, yes
}, 
%text=inside,
text=inside, 
]{43.5/{Manually, No}, 33.4/{Manually, Yes}, 12.9/{ML, No}, 6.0/{Both, Yes},1.7/{Both, No}, 2.6/{ML, Yes}}
\end{tikzpicture}
\caption{Share of innovations by matching method and matching status. Checked manually, using machine-learning (ML), or both (ML + manually). Matched to at least one patent (Yes) or matched to no patents (No). Innovations commercialized in 1970-2015}
\label{fig:treemap}
\end{figure*}
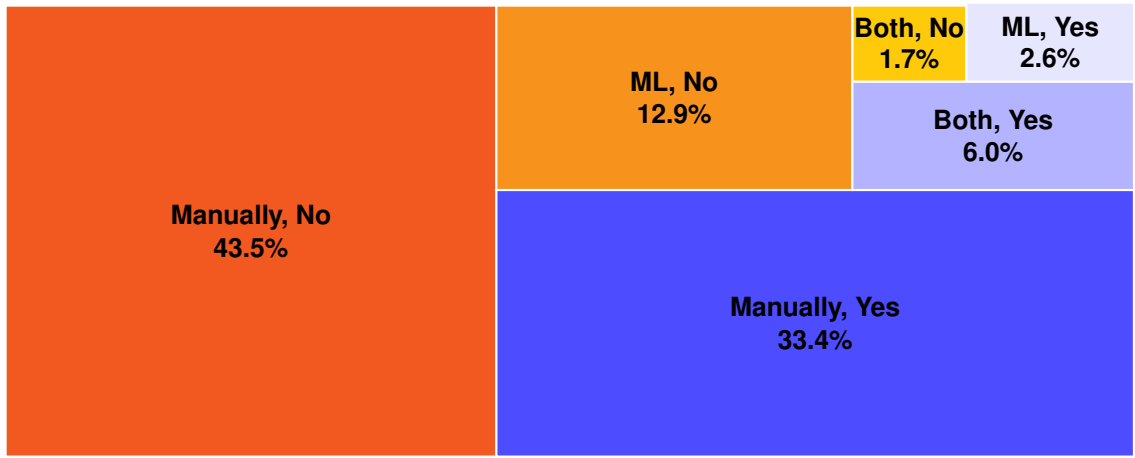

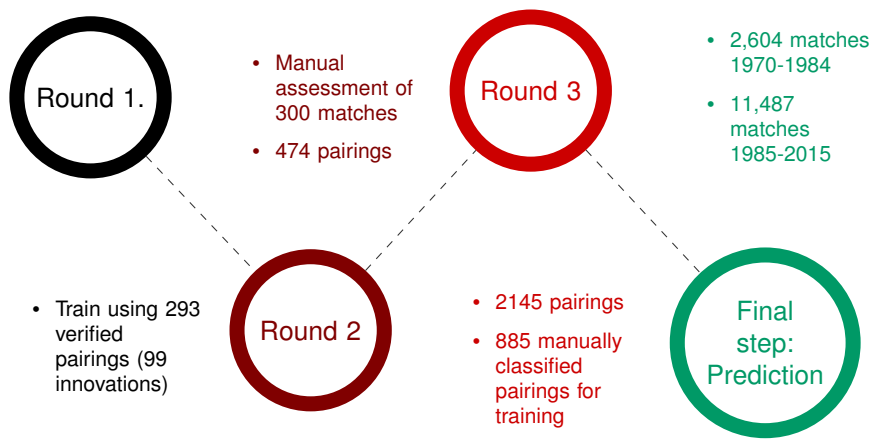
\begin{figure*}
\centering
\begin{tikzpicture}[
 bigcircle/.style={ 
    text width=1.6cm,
    align=center, 
    line width=2mm, 
    draw, 
    circle, 
    font=\sffamily\footnotesize 
  },
 desc/.style 2 args={ 
  text width=3cm, 
  font=\sffamily\scriptsize\RaggedRight, % set the font in the list
  label={[#1,yshift=-1.5ex,font=\sffamily\footnotesize]above:#2} % add the title as a label
  },
 node distance=10mm and 2mm 
]

\node [bigcircle] (circ1) {Round 1.};
\node [desc={black}{ },below=of circ1] (list1) {
\begin{itemize}
\setlength\itemsep{0pt} 
\item Train using 293 verified pairings (99 innovations)
%\item 
%\item Point 3
%\item Point 4
\end{itemize}
};

\node [bigcircle,black!50!red,right=of list1] (circ2) {Round 2};
\node [desc={black!50!red}{ },text=black!50!red, above=of circ2] (list2) {
\begin{itemize}
\setlength\itemsep{0pt}
\item Manual assessment of 300 matches
\item 474 pairings
%\item X,Y,Z 
%\item Point 4
\end{itemize}
};

\node [bigcircle,black!20!red,right=of list2] (circ3) {Round 3};
\node [desc={black!20!red}{},text=black!20!red, below=of circ3] (list3) {
\begin{itemize}
\setlength\itemsep{0pt}
\item 2145 pairings 
\item 885 manually classified pairings for training

\end{itemize}
};

\node [bigcircle,green!60!blue,right=of list3] (circ4) {Final step: Prediction };
\node [desc={green!60!blue}{},text=green!60!blue, above=of circ4] (h) {
\begin{itemize}
\setlength\itemsep{0pt}
\item 2,604 matches 1970-1984
\item 11,487 matches 1985-2015
\end{itemize}
};

\draw [dashed,black!80] (circ1) -- (circ2) -- (circ3) -- (circ4);
\end{tikzpicture} 
\caption{Overview of machine-learning process}
\label{fig:MLmatching}
\end{figure*}

\begin{figure*}[htbp]
\centering
\begin{subfigure}[b]{.45\linewidth}
\caption{}
\label{fig:machinecontrol}
\includegraphics[width=7cm, height=7cm, keepaspectratio]{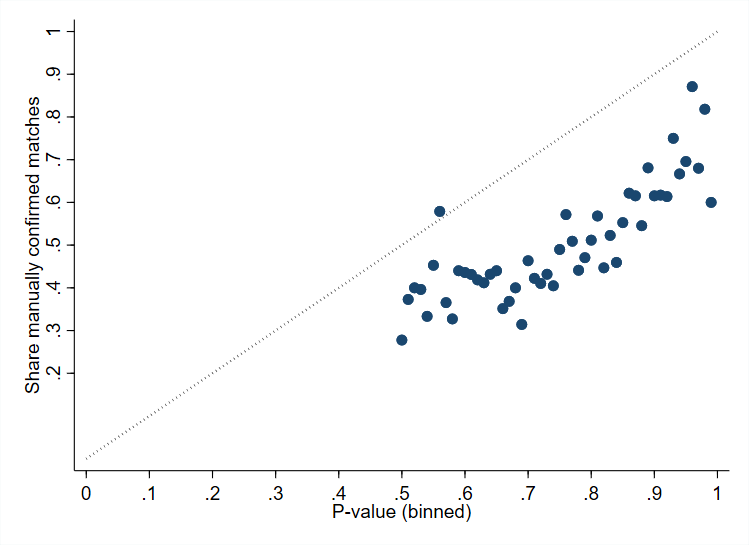}
\end{subfigure}
\centering
\begin{subfigure}[b]{.45\linewidth}
\centering
\caption{}
\label{fig:matchstatus}
\includegraphics[width=7cm, height=7cm, keepaspectratio]{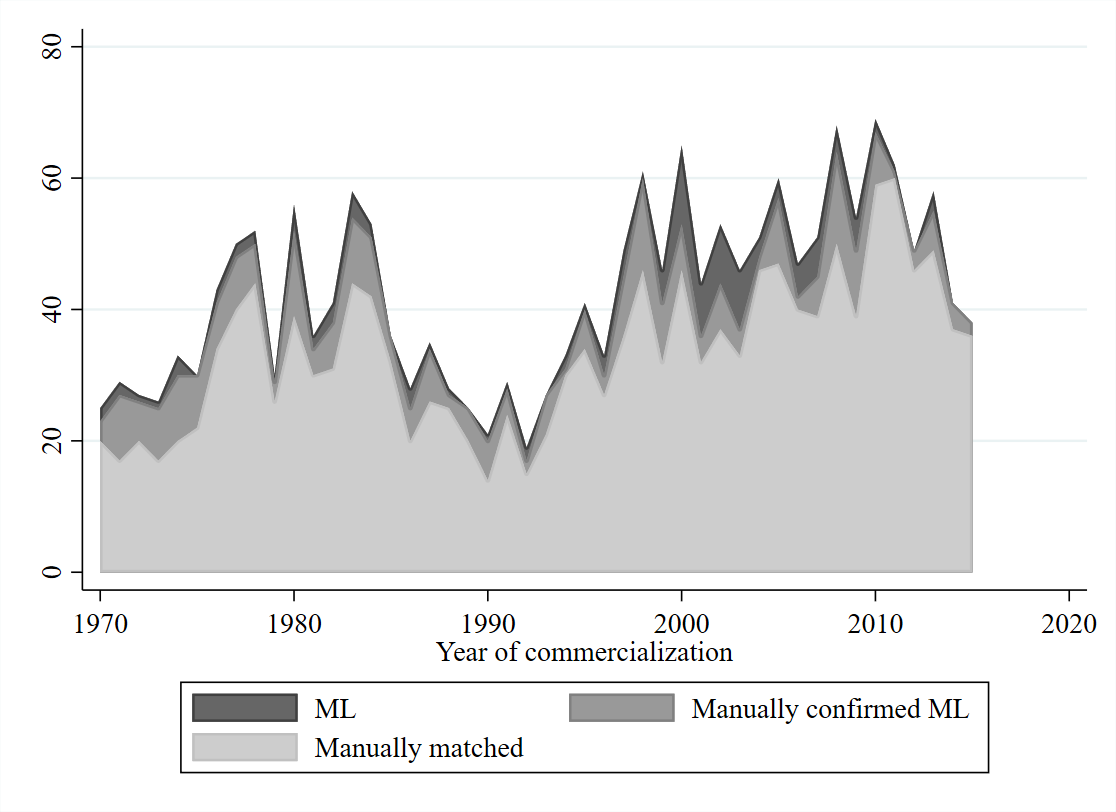}
\end{subfigure}
\begin{subfigure}[b]{.45\linewidth}
\centering
\caption{}
\label{fig:matchstatus_share}
\includegraphics[width=7cm, height=7cm, keepaspectratio]{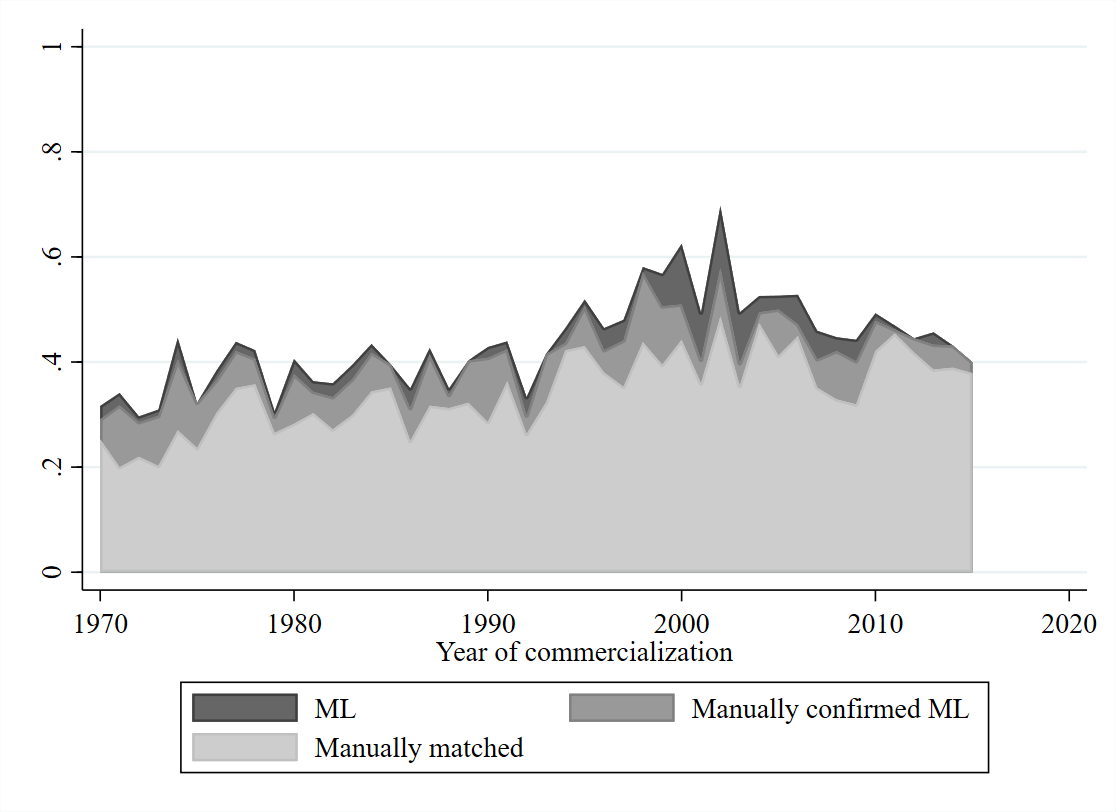}
\end{subfigure}
\begin{subfigure}[b]{.45\linewidth}
\centering
\caption{}
\label{fig:zipf}
\includegraphics[width=7cm, height=7cm, keepaspectratio]{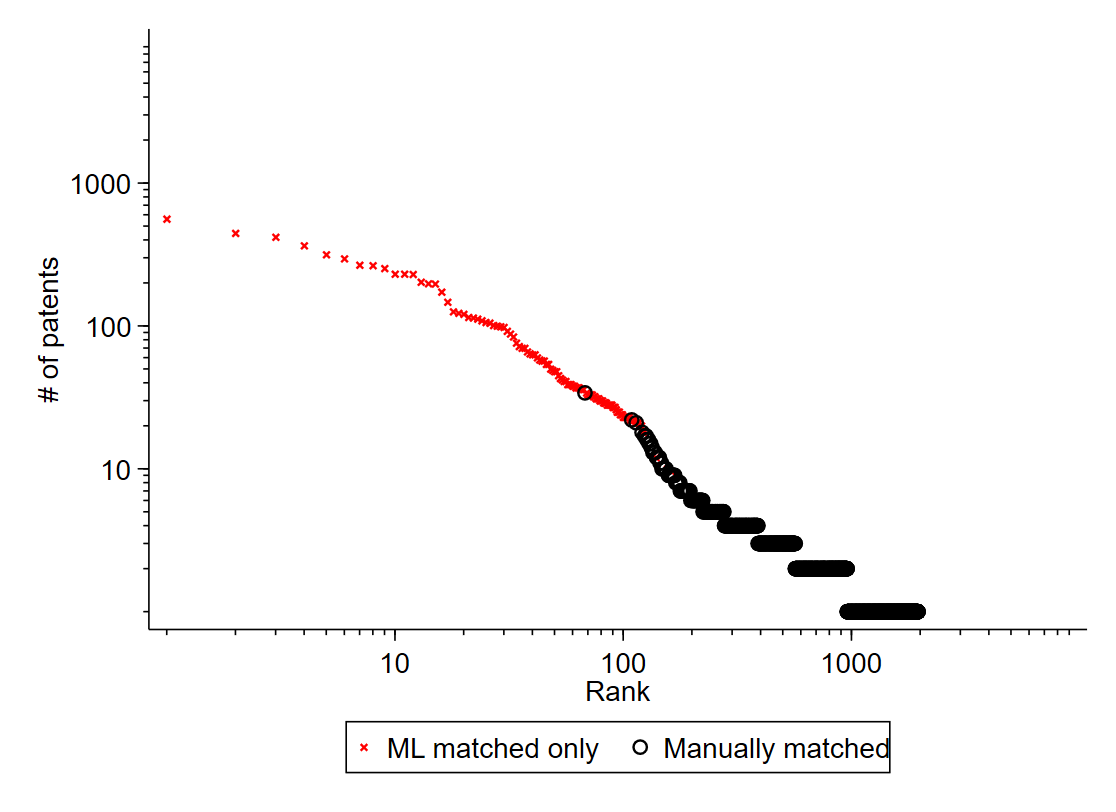}
\end{subfigure}
\caption{ (a) Manual validation. P-value from machine-learning (binned to two decimals) versus share of correct matchings in manual controls for 2,155 patent-innovation pairings, 1970-2015. Only suggested patent-innovation pairings with $P > 0.5$ were considered. (b) Number of patented innovations by commercialization year and matching method, (c) Share of innovations patented, by commercialization year and matching method. (d) Number of matched patents per innovation (y axis) and innovations ranking from innovation with most the patents to least (x axis) .}
\end{figure*}

\begin{figure*}[htbp]
\centering
\includegraphics[scale=0.4]{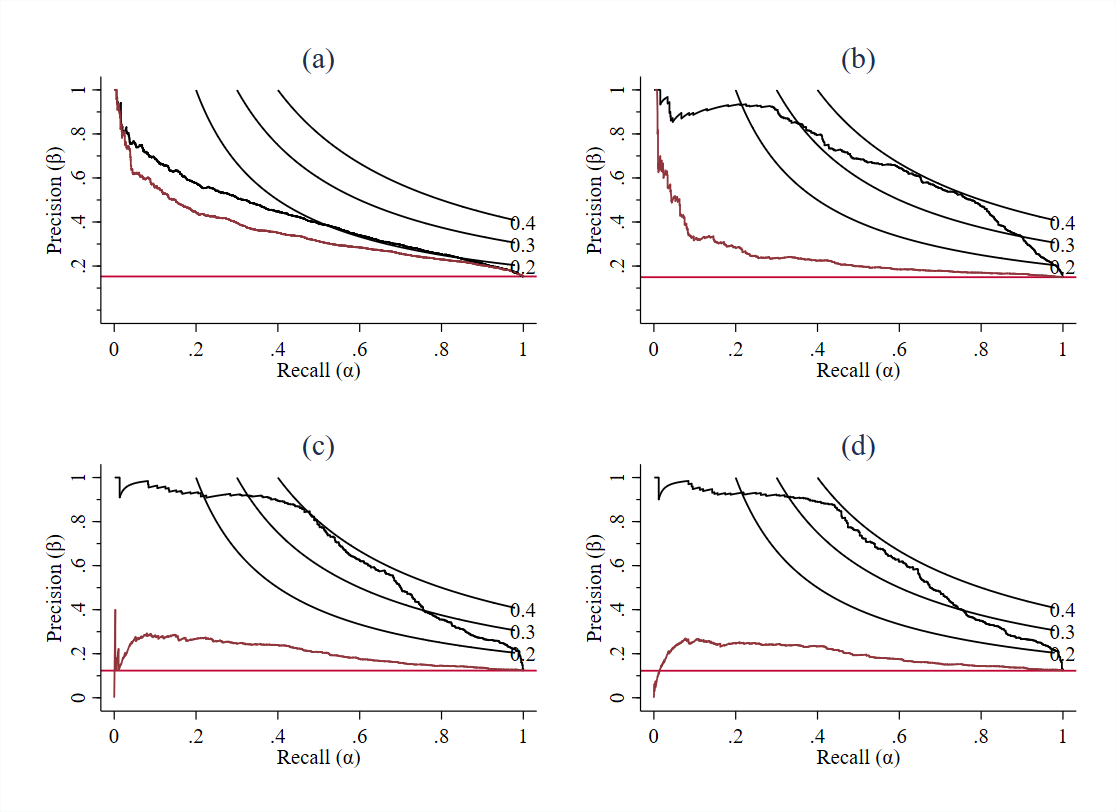}
\caption{(a) Precision and recall for prediction of LBIO (adjusted), (b) Blockbuster innovations, (c) new-to-the-world, and (d) principal components index. Predictions including dummies for technology field in black, without in red. The convex lines are points that achieve $\alpha \times \beta$ of 0.2, 0.3 and 0.4.}
\label{fig:precisionrecall_all}
\end{figure*}

However, manual matching is inadequate to ensure patent matching quality for big firms and highly complex innovations, such as systems consisting of many (patentable) parts. For such cases, the research team developed a machine-learning assisted methodology (step 2 in Fig. \ref{fig:matchingprocess}). This step departed from using keywords from trade journal articles, and subsequently trained a machine learning model to match keywords and other information to patent documents. The next step (step 3) was to perform manual checks on innovations for which the machine-learning model suggested less than 20 patents. The final database describes manually matched patents and patent-innovation matches suggested by machine-learning. The overall results are summarized in Fig. \ref{fig:treemap}. The manual matching resulted in 43.5\% of all innovations  classified as non-matches, and 33.4\% classified as positive matches. 12.9\% of all innovations could be concluded to be non-patented after the machine learning model failed to produce relevant patents. The rest of innovations were suggested by the machine-learning models to have a patent match. After screening most of these, another 6.0\% of all innovations could be confirmed to be patented, while 1.7\% were found to be false positives. The last two percent were innovations with more than 20 patents. These were not manually screened, but assumed to have at least one correct patent match.

The following sections give further detail about definitions and procedures to assure the quality of the manual and machine-learning assisted matching of innovations and patents.

\paragraph*{Manual matching}
The  matching process was based on trade journal articles and patent documents. To be considered a match, any  given patent had to meet the following three criteria:

\begin{enumerate}
\item The patent must be directly related to the innovation and/or the novel feature of the innovation, as described in a text
\item The patent document must contain a description (not just a title) linking the patent to the innovation
\item The patent must be filed within ten years before or after the commercialization year.  
\end{enumerate}

Trade journal articles were essential to determine the relevance of a certain patent.  These articles always, or in a majority of cases, provide detailed information about the innovation, including technical characteristics, the names of firm(s), inventors and personnel involved in the development of the innovation, the year the innovation was commercialization and biographical information, including whether the innovation was patented or had an associated patent application. 

The second and third criteria were enforced to ensure the quality of the matching. Patents lacking abstracts or descriptions were excluded since their relevance could not be decided from the title alone. 
In deciding what patents to match or rule out, the context of the patent abstract, description, and claim(s) were important. Technological background sections of patent documents could \emph{prima facie} be related to a particular technology field, but need to be directly related to the patent claim(s). Hence, such sections of patent documents were read with caution. 

Since technological knowledge ages fast, we also applied a window of ten years before or after the innovation's commercialization year. Although a few exceptions could be found, this window implies a broad definition of patent-innovation relationship, emphasizing the presence of technical links and allowing for well-known long development times of certain products, e.g., energy innovations and pharmaceuticals. Other studies have argued for smaller windows \cite{fontana2013SI}.

\paragraph*{Machine-learning assisted matching}
After the manual matching, the remaining challenge was to match innovations from big firms and innovations with high complexity to relevant patents. The patents of large firms, such as ABB, Scania, Volvo, and Ericsson, may range in the order of thousands or several thousands per year, which renders manual screening time consuming and impractical. 
For this reason, a subset of innovations were identified for which manual classification was deemed impractical or unreliable. This subset included large firms with over 200 granted patents, and more complex innovations that contain many patentable parts or require the combination of many development processes, examples of which are biotechnology, robotic systems, and airplane technology. For these innovations, the trade journal sources were read in order to assign a set of keywords. The keywords were selected in isolation from the machine-learning model and there was no upper boundary on the number of keywords that could be registered.

In the process of finding and matching patents for the above-mentioned innovations, three different subsets of innovations from the SWINNO database were used: first, a set of 99 innovations (commercialized between 1990 and 2015) with (manually) matched patents, a set of 645 innovations (1985-2015) that were used to further augment the model, and finally a set of 464 innovations (1970-1985). The machine-learning part of this process used scikit-learn \cite{scikit2011SI}.

To solve the machine learning problem, features needed to be constructed. An overview of the features used for machine learning is given in Table \ref{tab:features}. Apart from features involving information about the innovation's characteristics and its commercialization, the keyword information was used in several variables, in terms of the number and share of keywords that occurred in the patent's title, abstract and description. The process also made use of inventors and contact persons mentioned in trade journal articles.

\paragraph*{Finding viable patent candidates}
To find viable patent candidates, the Google Patents search engine was queried for each innovation with the firm name as well as the inventors (using all registered spelling variants from the SWINNO database) within 5 years of the year of commercialization. In the cases where more than 1,000 matches were found, the selection of used patents was reduced by constricting the year span until it either returned no more than 1,000 patents or the year span reached 0 (only searching for patents in the year of commercialization). After this limitation the set of 645 innovations for 1985-2015 led to 195,000 potential innovation-patent pairs. 

\paragraph*{Training models}

Models were trained on the annotated data in three rounds, as shown in Figure \ref{fig:MLmatching}, and further detailed elsewhere \cite{johansson2022SI}. The results from these rounds are summarized in Table \ref{tab:rounds} and Table \ref{tab:rounds-stats}. Initially, we used a sample of innovations matched manually to patents, of which we had the highest possible degree of confidence of a correct matching. For the second round, we let the model predict innovation-patent pairings from the 645 innovations from larger companies and went through the 300 most likely matches manually. With this additional data we trained a new model, and similarly processed 1,000 pairings, followed by 800 because model performance decreased with this additional data.  With this final dataset, we arrived at a model that we used to create short-lists of potential patents. 

\paragraph*{Manually processing the machine learning assisted matches}
To verify the robustness and reliability of the machine-learning-assisted matches, all innovations with less then 20 predicted patents were manually processed a final time. 
Manually processing the suggested pairings resulted in an assessment of whether the patent-innovation match, suggested by the machine-learning model, good be considered a true positive or a false positive. These patents were compared to the description of the innovation and how the constructed keywords were contextually used in the abstract, description, and claim(s), using the same methodology as in the previous manual matching procedure.

The results (Fig. \ref{fig:machinecontrol}) clearly suggest a positive correlation between the manual assessment and the machine-learning model's P value, viz. the model's estimated rate of false positives. The correlation between the fraction of verified positives and the machine-learning model's P value is shown in Fig.\ref{fig:machinecontrol}. The results also suggest that the machine-learning model tended to be slightly more confident in the matching than warranted. This overconfidence follows since the selection of potential patents was made to minimizes false negatives, whereas false positives were considered less problematic, as they can be dealt with by manual checks.

\paragraph*{Manually and machine-learning matched data}
\label{sec:results}

Figs. \ref{fig:matchstatus}-\ref{fig:matchstatus_share} show the number and share of patented innovations by commercialization year and matching method. The overall addition achieved by the machine-learning approach is only a minor share of the innovations (Fig. \ref{fig:treemap}), and the vast majority of the patented innovations were matched manually. However, the yearly fraction of innovations matched wholly or partially through machine learning could amount to some ten percentage points. 
Fig. \ref{fig:zipf} shows the number of patents matched to innovations versus the ranking of innovations from most patented to least, following the familiar Zipf's law pattern. Without machine-learning (the red observations) it would not be feasible to match the 100 innovations associated with the highest number of patents.

\paragraph*{Measure of information content}

The main question of interest to this study is how much information a patent analyst has about the members of the set of innovations. To measure this, the present study uses the expected number of patented innovations that a patent analyst can identify. This is measured as the product of patent propensity $\rho$, recall $\alpha$ and precision $\beta$ for reasons detailed below. 

The situation can be represented as follows. There are two sets: $X$ being the set of innovations, $Y$ being the set of patents. The intersection $X \cap Y$ is the set of patented innovations. The problem for the patent analyst is to tell which ones among the set of patents are innovations. Using patent quality data the patent analyst sets off to construct a model and makes a prediction of ``identified patents’’ $Z$. The set of true positives is the intersection $ Z \cap X$, and the share of true positives in the total potential innovations that could be defined is a measure of the quantity of information captured, the recall.

The number of true positives are not, due to the presence of false positives, identified with complete certainty. An arbitrary patent drawn will identify an innovation with probability $\vert Z \cap X \vert   / \vert Z \vert $, the precision. If the patent analyst is asked to give a list of the $\vert Z \cap X \vert $ true positives, on average  $\vert Z \cap X \vert \times  \vert Z \cap X \vert   / \vert Z \vert $ will be correctly identified. Expressed as a fraction of the total number of innovations, this, dimensionless, measure of information is
\begin{equation}
\vert Z \cap X \vert \times  \vert Z \cap X \vert   / \vert Z \vert \vert X \vert
\end{equation} which is the product of precision, recall and patent propensity
\begin{equation}
\rho \times \alpha \times \beta 
\end{equation}.

Expressed in general terms, this measure says how much correct information a receiver has about a particular message sent from a source, as a fraction of the total information in the source. 

One must not fail to notice that the standard in binary classification is to use the harmonic mean of precision and recall, known as $F_1$-score or its generalized variants ($F_\beta$), or sometimes the geometric mean, also known as Fowlkes-Mallow index. These indices quantify the trade-off between precision and recall, but do not offer a quantification of the information content. 

In standard information theory, the information obtained about one variable by observing another variable is typically measured in terms of mutual information \cite{shannon1948SI}. Put otherwise, this measures how much knowing one of these variables reduces the uncertainty, or entropy, of the other. 
The entropy of one variable is $H(X) = -\sum_{x \in X} p(X) \log p(X) $
The conditional entropy measures the amount of information needed to describe a variable $X$ given another one $Y$, $H(X \vert Y) = - \sum_{x \in X, y \in Y} p(X,Y) \log p(X,Y)/p(Y) $.
Mutual information is defined as $I(X,Y) = H(X)-H(X \vert Y) $ or $I(X,Y)= - \sum_{x \in X, y \in Y} p(X,Y) \log p(X,Y)/p(Y)p(X)$.

This measure, while theoretically attractive and widely used, has drawbacks in the current context. Firstly, mutual information does not distinguish between negative and positive associations between variables. Hence, the extreme cases $X=Y$ and $X=1-Y$ have the same mutual information.  Secondly, mutual information also takes into account true negatives. This is of no immediate interest in the current analysis.

\paragraph*{Logistic model}

To analyze the determinants of the propensity to patent an innovation $i$, logistic regression models were estimated, with dependent variable whether an innovation was patented (Y/N):

\begin{equation}
logit\left(E\left[P_{i}\mid X_{it}\right]\right) = \sum_k \alpha_k {comp}_{ik} + \sum_l \beta_l {nov}_{il} + \gamma {collab}_i + \delta_{s(i)} + \epsilon_{i}
 \end{equation} where ${comp}_{ik}$ is a set of variables measuring the complexity of the innovation, ${nov}_{il}$ a set of variable measuring its novelty, ${collab}_i$ measures whether the innovation has a collaboration, $\delta_{s(i)}$ are dummy variables for the sector $s$ of an innovation (ISIC Rev. 3, 2-digit level groups). We also use time dummies in all models (not shown in results). 
 
As a first measure of complexity, we distinguish between simple products and complex systems based on the description of the product innovation in the trade journal. We also classify innovations by the complexity of the knowledge base involved in developing the innovation, distinguishing between low complexity, medium complexity and high complexity. Low complexity involves only one major knowledge type, whereas high developmental complexity involves more than two types of knowledge.

The detailed information contained in trade journal articles also puts us in a position to assess the novelty of innovations. On the one hand, articles frequently state if the innovation implies a novelty to the world market. We also distinguish the novelty in relationship to the knowledge base of the firm. Radical innovations are those that have required a fundamental reorientation of the firm's knowledge base and/or those that were described in trade journal articles as a radical breakthrough from the perspective of the firm. A major improvement implies that the innovation meant a significant step, but did not require a radical reconfiguration of the firm's knowledge base. As a proxy for the significance of an innovation, we also use the number of times that an innovation was mentioned in distinct edited trade journal articles.

\paragraph*{Data on intellectual property rights}
To examine the role of generic developments for the propensity to patent in a given country, the analysis uses data on export shares to a given country (Statistics Sweden), and intellectual property rights \cite{ginarte1997SI,park2008SI}. The data on intellectual property rights measures five aspects of property rights protection quinquennially for 110 countries during the period 1960-2015. Each aspect is measured as an index between 0 and 1. Patent coverage measures the patentability of 7 types of inventions: software, pharmaceuticals, chemicals, food, plant and animal varieties, surgical products, micro-organisms and utility models \cite{park2008SI}.  
Patent duration is measured as a fraction of 20 years of patent from the date of application or, for grant-based patent systems, 17 years from the date of grant.

Patent membership measures whether a country partakes in international treaties, including the Paris convention and its revisions, the Patent cooperation treaty (PCT), Protection of new varieties (UPOV), the Budapest treaty (microorganism deposits) and the Trade-related intellectual property rights (TRIPS). 

The patent system's enforcement is scored according to whether it has (1) preliminary (pre-trial) injunctions, protecting patentees from infringement before trials, (2) contributory infringement, protecting against actions that do not in themselves infringe a patent,  or (3) burden-of-proof reversals, under which the burden of proof of non-infringement is placed on another party than the patentee, e.g., a company producing a product. 

The ``loss of rights'' variable measures the absence of three types of restrictions on patent rights: (1) working requirements, viz. requirements to put the patent into use to enjoy patent protection, (2) compulsory licensing, requiring patentees to share exploitation with third parties, and (3) revocation of patents for non-working. If there are no restrictions, the index takes value 1, and 0 if all are present \cite{ginarte1997SI,park2008SI}.

\paragraph*{Methods for analysis of patenting across countries}

In order to further our understanding of the determinants of patenting behavior, regressions were carried out to explain the propensity to patent an innovation $i$, commercialized in year $t$, in country $j$. 

Two sets of regressions were carried out. The appropriate approach depends on the properties of the outcome variable. Since the (binary) choice to patent an innovation in one country $j$ is neither mutually exclusive, nor generally independent from the choice to patent in another, ordinary logistic or probit regressions will be unsuitable for statistical inference. Instead, a multivariate probit approach was used to account for correlation of the binary outcomes. 

The propensity of an innovation $i$ to be patented in country $j$ was analyzed through:

\begin{equation}
y_{ij}^{*} = \beta_{j}^{'} X_{ij} + \epsilon_{ij}
\end{equation} with $y_{ij} = 1$ if $y_{ij}^{*} > 0$, otherwise $0$.

Specifically, the model used is
\begin{equation}
y_{ij}^{*}= \sum_k \alpha_k {comp}_{ik} + \sum_l \beta_l {nov}_{il} + \gamma {collab}_i + \delta_{s(i)} + \sum_m {IPR}_{jmt} +X_{jt}+\epsilon_{ijt}
 \end{equation} where ${IPR}_{jmt}$ are the set of variables that capture aspects of patent laws in country $j$ and year $t$, as detailed above. ${comp}_{ik}$ is a set of variables measuring the complexity of the innovation, ${nov}_{il}$ a set of variable measuring its novelty, ${collab}_i$ measures whether the innovation has a collaboration, $\delta_{s(i)}$ are dummy variables for the sector $s$ of an innovation (ISIC Rev. 3, 2-digit level groups). 
 
\paragraph*{Decomposition}
To understand whether the changes in patenting propensity was driven by generic, sectoral or structural changes it is possible to carry out a straightforward decomposition of the aggregate patent propensity. If we define the average patent propensity $\bar{p}$ as $\bar{p}=\sum_i w_i p_i$ with $w_i$ the share of innovations of a sector $i$ and $p_i$ the patent propensity in sector $i$, we can easily derive the percentage point change as the sum of three components.

\begin{equation}
\Delta \bar{p} \equiv \underbrace{\sum_i \Delta w_i p_i}_{{between}} + \underbrace{\sum_i \Delta p_i w_i}_{{within}} + \underbrace{\sum_i \Delta p_i \Delta w_i }_{{interaction}}
\end{equation}

Apart from the within and between effects, the ``interaction effect'' captures the effect that changes in patent propensity might be (positively or negatively) correlated with changes in sectoral shares (compare the celebrated Price equation; see \cite{frank1997SI}).

\subsection*{Supplementary text}

\paragraph*{Determinants of the propensity to patent}
 To gain insight into what patent indicators capture, and fail to capture, logistic regressions were applied to analyze the determinants of whether an innovation $i$ has a patent. In a first set of models, the models used variables on innovation characteristics, sectoral dummies, and the role of intellectual property rights, as detailed in materials and methods.

The results are shown in Table \ref{tab:logistic}. The baseline model examines the impact of the variables on the overall patent propensity. The first panel includes only sectoral dummies, showing overall a similar picture as in Table \ref{tab:sectoral}. The second panel tested for the impact of complexity, showing that patented innovations tend to not be complex systems, but tend, on the other hand, to have higher developmental complexity. This may seem contradictory, but rather reflects different preconditions and means of appropriation. On the one hand, complex system innovations may be more difficult and costly to copy, there may be significant barriers to entry, and lead time advantages are likely to be a more efficient means of appropriation \cite{arundel1998SI}. On the other hand, high-tech industries such as pharmaceuticals, machinery and electronic engineering are known to have lost costs of copying innovations, and therefore high incentives for securing returns from innovation \cite{arundel1998SI, breschi2000SI}. 

The third panel included measures of novelty. The results unanimously suggest that patented innovations are associated with all kinds of novelty and significance. We observe significant and robust positive associations with the variables measuring whether the innovation is new to the world, radical or major improvement, as well as the number of sources an innovation, thought to be a proxy for significance. Innovations that result from collaborations between many firms are, overall, less likely to be patented, but the finding is less robust.

Panels 4-6 use as dependent variable whether the innovation has a patent in EPO, USPTO or Sweden. The results are overall very similar and hence robust across different patent indicators. 

Panels 7 and 8 run complementary regressions. Panel 7 runs a negative binomial count model regression to predict the number of patents an innovation has, with similar results, except that complex systems are now insignificant. Panel 8 predicts the number of countries in which the innovation is patented, again using a negative binomial count model. Again, the results are very similar to the other specifications. Notably, the number of patents an innovation has, and the number of countries in which an innovation has patents are positively correlated to our measures of novelty. This is in line with the notion that patent family size captures significance \cite{lanjouw2004SI}.
 
\paragraph*{Determinants of patenting across countries}

Since the multivariate probit model is computationally intensive, the analysis focused on the five most common national patent offices: Sweden, US, Germany, Japan and Canada. Together these five patent offices account for virtually all patented innovations (98.7\%, compare Fig. \ref{fig:patprop_cum}). For completeness, Table \ref{tab:countrylevel} also reports ordinary logistic models for a panel of all countries with available IPR data.

Table \ref{tab:mvprobit} and Table \ref{tab:countrylevel} show overall similar results, but the multivariate probit regressions suggest the presence of cross-country heterogeneity in the effects of IPR on patent propensity. Membership in international treaties (e.g., PCT and TRIPS) has the most consistent positive effect for patent propensity. The overall effect of duration in (Table \ref{tab:countrylevel}) is positive, but of the five focal countries, only Sweden and Japan saw changes in patent duration (Table \ref{tab:mvprobit}). The effect of patent coverage of seven types of inventions (software, pharmaceuticals, chemicals, food, plant and animal varieties) is less consistent, both in the multivariate probit and logistic regressions.  Loss of rights, measuring the absence of restrictions, e.g., compulsory licensing, on patent rights, has a negative impact on patent propensity in general, but a positive coefficient for Sweden. The negative signs observed for patent coverage and loss of rights, could reflect an anticommons effect \cite{heller1998SI}. 

Using these numbers, it is possible to calculate an upper bound of the importance of IPR for innovation activity. A somewhat generous interpretation of the results is that a patented innovation that depends greatly on changes in IPR would not have been forthcoming without them. This is ``generous'', because a great importance of IPR does not necessarily imply that patents were immediately important to the development or commercialization of the innovation, but may instead imply that innovators have found it of greater importance to employ defensive patents, due to a generally increased use of patents \cite{blind2009SI}. For this reason, the coefficients should be viewed as estimating an \emph{upper bound} of the importance of IPR for innovation activity. Conversely, innovations that were not patented or whose patenting choice does not depend on IPR ought to be viewed as relatively independent of policy changes. 

To illustrate the magnitude of the importance of IPR, one may assume that the internationally strengthened IPR in the 1970s, 1980s and 1990s never happened and estimate counterfactual rates of patenting for each innovation and country, using the IPR indices for 1970. Using the multivariate probit model, the probability than an innovation had at least one patent was estimated for the ordinary regression and a regression with counterfactual 1970 levels of IPR (coverage, membership, loss of rights, patent duration and exclusion). The number of ``lost'' innovations is the difference in the expected number of patented innovations with the historically accurate IPR levels and the counterfactual. The results are shown in Fig. \ref{fig:counterfactual}. Under the above interpretation, an estimated 8\% of all innovations, or would not have been forthcoming without the strengthened patent laws. After the TRIPS agreement, signed in 1994, this figure was 9.8\% , reaching a peak of 15.5\% in the late 1990s.

As an indication, Fig. \ref{fig:counterfactual} also shows the share of patents whose filing date precede the commercialization date. One may reason that patents filed after commercialization were either strategic, or less important for the development of the innovation \cite{blind2009SI}. The results do not indicate an increase of post-commercialization patenting over the period, as could be expected if patenting activity had largely shifted towards an predominantly defensive strategies.

\paragraph*{Predicting innovations from patent data}

To investigate the ability of patent quality data to predict significant innovations, this study uses the EPO patents extracted from Google Patents and matched to the OECD Patent Quality Indicators database (January 2021; \cite{squicciarini2013SI}). 

The analysis departs from classifying patents into (1) those linked to significant innovations and (2) a set of non-significant patents. Since no such classification is beyond criticism, we test the discriminatory capacity of patent quality measures on four classifications. 

A basic comparison would be between all patents linked to the LBIO innovation database, and patents that have no such link. A problem with this comparison, is, however, that  it is plausible that there are significant patents among those not linked to the LBIO innovation database. For this reason, patents with more than 5 years of renewal were excluded from the analysis and compare LBIO innovations with  those patents that were only renewed for 5 years or less.

As the benchmark, the analysis makes us of a set of 40 ``blockbuster innovations'', widely known to have had been major success stories and having had a major economic impact. These are pooled together from written sources and interviews with major innovating firms. A first source lists 100 major Swedish patented innovations from 1945 to 1980, selected on the basis of having generated a turnover of at least 3.5 million Swedish krona in 1980 prices \cite{wallmark1991SI}. Another source was an extended list of major Swedish innovations during the 20th century up until 2002 \cite{sedig2002SI} and identify more recent major successes including Skype, Spotify and the bicycle helmet Hövding. In addition, interviews were carried out with research directors in major companies to separate highly successful from incremental or unimportant products. The companies are Ericsson, AGA, Atlas Copco and Sandviken. These innovations correspond to 923 EPO patents.

Two other definitions of significant innovations are also used as robustness controls: \begin{itemize}
\item ``New to the world'', 795 patents linked to LBIO innovations, described as new on the world market and mentioned in at least 3 journal articles, excluding innovations that are incremental from the firm perspective. 
\item ``PCA'', selection of  807 patents. An index (principal components) based on the number of trade journal sources of an innovation, its market novelty and firm novelty.
\end{itemize}

The last subset is constructed from principal components analysis (PCA) on the variables market novelty, firm novelty and the number of sources in which the innovation is mentioned to construct an index of overall significance. The innovations included are those that score in the highest decile.

Incremental patents, are identified by using trade journal information and interviews. Non-significant innovations are innovations that were described in trade journal articles as incremental from the firms' perspective, viz. a product improvement, were not indicated as new to the Swedish or world market, and were only mentioned in one trade-journal source. This set also includes innovations that were confirmed by interviewees to have been of lesser or no importance.

All selections of significant innovations are compared against the identified incremental patents. 
An issue is that the performance of the regressions, including the precision and recall, is influenced by the ratio of positives (significant patents) to negatives (incremental patents). One way of approaching this issue, would be to assume that the ratio of positive to negative outcomes is the same as the ratio of LBIO to Non-LBIO patents. The fraction of LBIO patents is 4.8\%. This assumption however risks underestimating the performance since there may be significant patents among the Non-LBIO patents. Instead the approach is to identify the most significant innovations among the LBIO data and estimate upper bounds. The criteria of the above selections are chosen so as to achieve a fraction of the ca 8-9\% most significant patents in the LBIO data, and the ratio of positive to negative outcomes in the models are construed such that \emph{at least} as many of the observations are positive outcomes. In practice, the fraction of positive outcomes are in above 10\%, implying that the results are possibly upward biased. 

As predictive variables, the regressions use the number of patent citations (forward citations) within 7 years from the publication date, the number of years for which a patent was renewed, and the patent family size \cite{squicciarini2013SI}. In addition, the regressions include two indices. The originality of a patent, captures the breadth of knowledge (technology fields) that a patent relies on. The originality of a patent $p$ is calculated as \cite{squicciarini2013SI}

\begin{equation}
{Originality}_p = 1 - \sum_{j}^{np} s_{pj}^2
\end{equation} where $s_{pj}$ is the share of citations made by patent $p$ to patent class $j$ out of the $n_p$ patent codes contained in the patents cited by patent $p$.

The analysis also makes use of an index of radicalness \cite{squicciarini2013SI} to capture the diversity in the technologies that a patent relies upon, calculated as

\begin{equation}
{Radicalness}_p = \sum_{j}^{np} {CT}_{j}/n_p
\end{equation} with $CT_j$ the number of technology classes of a patent $j$ cited by patent $p$ and $n_p$ is the number of patent codes contained in the patents cited by patent $p$.

Descriptive statistics support differences in the number of patent citations received depending on whether the patent was connected to the LBIO, a blockbuster patent, new to the world or the PCA index (Table \ref{tab:citations}). Patents not connected to the LBIO database have a mean number of citations of 0.76. LBIO innovations have on average more than double the number of citations (1.57), while the blockbuster innovations, selected as being the most successful Swedish innovations, have, on average, only somewhat higher number of citations than the average LBIO innovation (1.82). 

Figs. \ref{fig:PCscatter} and \ref{fig:stripplot_pc1} use principal components to compare the overall properties of the non-LBIO, LBIO and blockbuster samples. These are based on citations (log) when technology fields and filing dates are controlled for, originality, radicalness, renewal and family size (see Table \ref{tab:pca}). 

Regressions are run with and without dummies for the patents' technology field (Tables \ref{tab:predictsuper_without} and \ref{tab:predictsuper} respectively). The results suggest first of all that LBIO innovations (model 1) are, again, linked to several quality measures. High number of forward citations, high originality, patent family size and patent renewal are all positively associated with LBIO innovations. Models 2-5 compare patents linked to significant innovations with non-significant patents as outlined above. 

From these models it is clear that none of the patent quality measures are, on their own, a consistent predictor of significant innovations. The results vary between the definitions used to identify significance. The LBIO data has expected (positive) signs in all coefficients, but for the blockbuster benchmark only family size has a significant positive effect. The most consistent predictor is patent renewal, although blockbuster patents have no significant positive effect. 

The predictive power of the model may also be evaluated in terms of the precision and recall, shown for Models 2-5 in Figure \ref{fig:precisionrecall_all}. The best performing model is the one predicting blockbuster innovations, achieving a maximum recall $\alpha$ of 0.561 and precision $\beta$ of 0.7, when controls for technology field are included. This results in a maximum product of 0.393. Similar results are achieved for the models predicting new-to-the-world innovations and the principal components. 
 
\clearpage

\subsection*{Supplementary Figures}

\clearpage
\subsection*{Supplementary tables}

\begin{table*}[htbp]
\centering \caption{Summary statistics of patent propensity, by commercialization year, 1970-1989 (top panel), 1990-2015 (second panel) and total (bottom panel)
\label{tab:sumstat}}
\footnotesize
\begin{tabular}{l c c  c}\hline\hline
\multicolumn{1}{c}{\textbf{Variable}} & \textbf{Mean}
 & \textbf{Std. Dev.} & \textbf{N}\\ \hline
\emph{1970-1989} &  &  & \\
Patented (Y/N) & 0.377 & 0.485  & 1968\\
EPO$^{***}$ & 0.181 & 0.385  & 1350\\
USPTO & 0.266 & 0.442  & 1968\\
SE & 0.308 & 0.462  & 1968\\
JPO & 0.195 & 0.396  & 1968\\
\hline

\emph{1990-2015} &  &  & \\
Patented (Y/N) & 0.488 & 0.5  & 2492\\
EPO & 0.414 & 0.493  & 2492\\
USPTO & 0.378 & 0.485  & 2492\\
SE & 0.353 & 0.478  & 2492\\
JPO & 0.255 & 0.436  & 2492\\
\hline
\emph{1970-2015} &  &  & \\
Patented (Y/N) & 0.439 & 0.496  & 4460\\
EPO$^{***}$ & 0.332 & 0.471  & 3842\\
USPTO & 0.328 & 0.47  & 4460\\
SE & 0.333 & 0.471  & 4460\\
JPO & 0.229 & 0.42  & 4460\\
\hline
\end{tabular}
\caption*{\footnotesize{
$^{***}$ EPO was established in 1977. The figures therefore are based on innovations commercialized 1977-2015.
}
}
\end{table*}

\begin{table*}[hptb]
\footnotesize
\centering
\begin{tabular}{lHccccH} \hline
& & 1970-1989 & & 1990-2015 & & \\ 
Sector & period1 & Innovations & Share patented & Innovations & Share patented & \\ 
\hline
Foodstuff & 1 & 11 & 0.273 & 36 & 0.417 & 12 \\
Textiles & 1 & 11 & 0.545 & 17 & 0.471 & 6 \\
Wood & 1 & 27 & 0.296 & 56 & 0.393 & 12 \\
Pulp \& paper & 1 & 19 & 0.211 & 45 & 0.489 & 10 \\
Publishing & 1 & - & 0.500 & 6 & 0.333 & 12 \\
Petroleum & 1 & - & - & - & - & 12 \\
Chemicals & 1 & 50 & 0.420 & 80 & 0.600 & 12 \\
Pharmaceuticals & 1 & 18 & 0.556 & 52 & 0.846 & 11 \\
Plastics & 1 & 100 & 0.420 & 115 & 0.504 & 12 \\
Other non-metallical & 1 & 27 & 0.481 & 14 & 0.357 & 7 \\
Basic metals & 1 & 49 & 0.469 & 54 & 0.519 & 3 \\
Fabricated metals & 1 & 121 & 0.455 & 118 & 0.449 & 12 \\
Machinery & 1 & 657 & 0.451 & 486 & 0.525 & 12 \\
Computers & 1 & 140 & 0.214 & 81 & 0.346 & 12 \\
Electrical app. & 1 & 80 & 0.338 & 102 & 0.549 & 1 \\
Telecom. eq. & 1 & 88 & 0.227 & 244 & 0.492 & 2 \\
Electronic eq. & 1 & 272 & 0.316 & 402 & 0.557 & 10 \\
Automotive & 1 & 68 & 0.368 & 73 & 0.438 & 12 \\
Other transp. eq. & 1 & 58 & 0.190 & 41 & 0.244 & 12 \\
Other manufacturing & 1 & 16 & 0.188 & 27 & 0.444 & 12 \\
Recycling & 1 & 12 & 0.583 & 13 & 0.462 & 12 \\
Telecommunication services & 1 & - & - & 16 & 0.625 & 12 \\
Software & 1 & 40 & 0.0750 & 295 & 0.322 & 3 \\
R\&D & 1 & - & - & 36 & 0.806 & 12 \\
Other business services & 1 & 75 & 0.533 & 60 & 0.383 & 5 \\
 \hline
\end{tabular}
\caption{Patent propensity across sectors (ISIC Rev. 3), all patent offices. \footnotesize{Note: Results not given when sectoral counts are below five.}}
\label{tab:sectoral}
\end{table*}

\begin{table}[hptb]
\footnotesize
\centering
\begin{longtable*}{p{2,5cm}ccccccHcHcH} \hline
 & (1) & (2) & (3) & (4) & (5) & (6) & (7) & (7) & (7) & (8) & (8) \\
VARIABLES & Baseline & Complexity & Novelty & EPO & USPTO & SE & WO & \# patents & \# patents & \# countries & \# countries \\ \hline
 &  &  &  &  &  &  &  &  &  &  &  \\
Simple product &  & 0.0366 & -0.0235 & -0.0540 & -0.153 & -0.160 & -0.0105 & -0.0364 &  & 0.176 &  \\
 &  & (0.104) & (0.108) & (0.126) & (0.115) & (0.112) & (0.131) & (0.0794) &  & (0.116) &  \\
Complex system &  & -0.321*** & -0.295*** & -0.0673 & -0.322*** & -0.199** & -0.0429 & -0.119* &  & -0.188* &  \\
 &  & (0.0922) & (0.0949) & (0.111) & (0.101) & (0.0984) & (0.116) & (0.0701) &  & (0.102) &  \\
Low dev. complexity &  & -0.226** & -0.132 & -0.294** & -0.170 & -0.0496 & -0.255* & -0.0873 &  & -0.242** &  \\
 &  & (0.0964) & (0.0992) & (0.125) & (0.109) & (0.105) & (0.130) & (0.0789) &  & (0.109) &  \\
High dev. complexity &  & 0.520*** & 0.317*** & 0.356*** & 0.505*** & 0.324*** & 0.406*** & 0.271*** &  & 0.244** &  \\
 &  & (0.0941) & (0.0981) & (0.108) & (0.0995) & (0.0986) & (0.111) & (0.0670) &  & (0.103) &  \\
New to the world &  &  & 0.161* & 0.0872 & 0.332*** & 0.136 & 0.0774 & 0.157*** &  & 0.263*** &  \\
 &  &  & (0.0836) & (0.0948) & (0.0860) & (0.0849) & (0.0988) & (0.0595) &  & (0.0929) &  \\
Radical &  &  & 0.745*** & 0.662*** & 0.692*** & 0.643*** & 0.872*** & 0.716*** &  & 0.709*** &  \\
 &  &  & (0.114) & (0.139) & (0.126) & (0.121) & (0.146) & (0.0929) &  & (0.125) &  \\
Major improvement &  &  & 0.233** & 0.164 & 0.245** & 0.266** & 0.213 & 0.271*** &  & 0.367*** &  \\
 &  &  & (0.109) & (0.137) & (0.122) & (0.117) & (0.144) & (0.0916) &  & (0.118) &  \\
Sources &  &  & 0.349*** & 0.304*** & 0.313*** & 0.208*** & 0.327*** & 0.160*** &  & 0.143*** &  \\
 &  &  & (0.0367) & (0.0355) & (0.0335) & (0.0293) & (0.0370) & (0.0173) &  & (0.0329) &  \\
Collaboration &  & -0.0656 & -0.184** & -0.161* & -0.287*** & -0.199** & -0.132 & -0.129** &  & -0.216** &  \\
 &  & (0.0798) & (0.0832) & (0.0969) & (0.0897) & (0.0868) & (0.100) & (0.0625) &  & (0.0909) &  \\
Textiles & 0.796** & 0.504 & 0.677* & 0.636 & 0.857** & 0.357 & 0.315 & 0.483* &  & 0.280 &  \\
 & (0.377) & (0.371) & (0.385) & (0.458) & (0.391) & (0.398) & (0.492) & (0.280) &  & (0.418) &  \\
%Wood & 0.0498 & -0.0548 & -0.115 & -0.158 & -0.599* & -0.176 & -0.0112 & -0.169 &  & -0.420 &  \\
% & (0.272) & (0.264) & (0.274) & (0.326) & (0.330) & (0.287) & (0.324) & (0.215) &  & (0.302) &  \\
%Pulp \& paper & 0.313 & 0.116 & -0.00767 & 0.292 & -0.0650 & 0.324 & 0.317 & 0.222 &  & -0.174 &  \\
% & (0.295) & (0.286) & (0.299) & (0.336) & (0.321) & (0.299) & (0.347) & (0.211) &  & (0.322) &  \\
Chemicals & 0.875*** & 0.525** & 0.515** & 0.569** & 0.456* & 0.158 & 0.536** & 0.259 &  & 0.217 &  \\
 & (0.227) & (0.217) & (0.226) & (0.256) & (0.234) & (0.233) & (0.266) & (0.164) &  & (0.241) &  \\
Pharmaceuticals & 1.906*** & 1.393*** & 1.210*** & 1.501*** & 1.388*** & 0.0179 & 1.425*** & 1.063*** &  & 0.885*** &  \\
 & (0.321) & (0.313) & (0.320) & (0.325) & (0.313) & (0.285) & (0.327) & (0.172) &  & (0.301) &  \\
Plastics \& rubber & 0.658*** & 0.463** & 0.507*** & 0.314 & -0.127 & 0.274 & 0.298 & 0.103 &  & 0.130 &  \\
 & (0.197) & (0.180) & (0.185) & (0.222) & (0.208) & (0.194) & (0.231) & (0.145) &  & (0.200) &  \\
Basic metals & 0.722*** & 0.481** & 0.579** & 0.654** & 0.592** & 0.297 & 0.284 & 0.227 &  & 0.346 &  \\
 & (0.245) & (0.234) & (0.242) & (0.285) & (0.249) & (0.252) & (0.303) & (0.188) &  & (0.262) &  \\
Fabricated metals & 0.567*** & 0.385** & 0.453*** & 0.390* & 0.106 & 0.394** & 0.289 & 0.189 &  & 0.223 &  \\
 & (0.192) & (0.171) & (0.174) & (0.209) & (0.189) & (0.179) & (0.220) & (0.137) &  & (0.192) &  \\
Machinery & 0.747*** & 0.562*** & 0.623*** & 0.399*** & 0.324** & 0.485*** & 0.432*** & 0.297*** &  & 0.261* &  \\
 & (0.151) & (0.123) & (0.126) & (0.151) & (0.134) & (0.130) & (0.158) & (0.0984) &  & (0.140) &  \\
Computers & -0.266 & -0.520*** & -0.503** & -0.576** & -0.670*** & -0.764*** & -0.407 & -0.331** &  & -0.850*** &  \\
 & (0.210) & (0.191) & (0.197) & (0.246) & (0.219) & (0.211) & (0.255) & (0.156) &  & (0.207) &  \\
Electrical app. & 0.570*** & 0.317* & 0.216 & 0.327 & -0.0504 & -0.0839 & 0.379 & 0.162 &  & -0.0929 &  \\
 & (0.206) & (0.187) & (0.193) & (0.227) & (0.206) & (0.202) & (0.236) & (0.143) &  & (0.211) &  \\
%Telecom eq. & 0.281 & -0.0267 & -0.0292 & 0.00196 & -0.0913 & -0.350** & 0.0441 & -0.200 &  & -0.266 &  \\
% & (0.180) & (0.158) & (0.162) & (0.182) & (0.171) & (0.171) & (0.189) & (0.125) &  & (0.179) &  \\
Electronic eq. & 0.543*** & 0.247* & 0.223 & 0.285* & 0.116 & 0.0213 & 0.356** & 0.113 &  & -0.200 &  \\
 & (0.160) & (0.133) & (0.137) & (0.159) & (0.145) & (0.142) & (0.166) & (0.105) &  & (0.151) &  \\
%Automotive & 0.325 & 0.132 & 0.279 & 0.151 & 0.238 & 0.108 & 0.125 & 0.0298 &  & 0.0234 &  \\
% & (0.225) & (0.206) & (0.210) & (0.254) & (0.223) & (0.218) & (0.267) & (0.170) &  & (0.233) &  \\
%Recycling & 0.836* & 0.585 & 0.482 & 0.0133 & 0.558 & -0.112 & 0.0265 & 0.0357 &  & 0.0528 &  \\
% & (0.430) & (0.424) & (0.442) & (0.572) & (0.450) & (0.473) & (0.584) & (0.322) &  & (0.487) &  \\
Software & -0.388** & -0.643*** & -0.750*** & -0.916*** & -0.600*** & -1.323*** & -0.756*** & -0.529*** &  & -1.106*** &  \\
 & (0.187) & (0.166) & (0.172) & (0.196) & (0.181) & (0.201) & (0.197) & (0.131) &  & (0.183) &  \\
R\&D & 2.104*** & 1.719*** & 1.471*** & 1.788*** & 1.757*** & -0.123 & 1.547*** & 0.937*** &  & 0.565 &  \\
 & (0.447) & (0.440) & (0.450) & (0.455) & (0.454) & (0.381) & (0.443) & (0.218) &  & (0.396) &  \\
%lnalpha &  &  &  &  &  &  &  &  & 0.000255 &  & 1.589*** \\
% &  &  &  &  &  &  &  &  & (0.0617) &  & (0.0327) \\
Constant & -1.397*** & -1.144*** & -1.886*** & -3.986*** & -1.823*** & -2.569*** & -5.629*** & -1.633*** &  & -0.0821 &  \\
 & (0.282) & (0.272) & (0.297) & (0.493) & (0.315) & (0.367) & (1.038) & (0.244) &  & (0.308) &  \\
Observations & 4,460 & 4,460 & 4,460 & 3,842 & 4,460 & 4,460 & 4,126 & 4,460 & 4,460 & 4,460 & 4,460 \\
 R-squared & 0.0485 & 0.0541 & 0.0938 & 0.150 & 0.105 & 0.0844 & 0.217 & 0.0768 & 0.0768 & 0.0206 & 0.0206 \\ \hline
\multicolumn{12}{c}{ Standard errors in parentheses} \\
\multicolumn{12}{c}{ *** p$<$0.01, ** p$<$0.05, * p$<$0.1} \\
\end{longtable*}
\caption{Logistic regression (log odds ratios). Dependent variables as described in text. Selected sectors shown.}
\label{tab:logistic}
\end{table}

\begin{table*}[hptb]
\footnotesize
\centering
\begin{tabular}{lcccccHHHHHHHHHH} \hline
 & (1) & (2) & (3) & (4) & (5) & (6) & (7) & (8) & (9) & (10) & (11) & (12) & (13) & (14) & (15) \\
VARIABLES & Sweden & US & Germany & Japan & Canada & Baseline & Baseline & Baseline & Baseline & Baseline & Baseline & Baseline & Baseline & Baseline & Baseline \\ \hline
%paste here
Simple product & -0.0662 & -0.0652 & -0.0191 & -0.0575 & -0.0159 &  &  &  &  &  &  &  &  &  &  \\
 & (0.0651) & (0.0660) & (0.0674) & (0.0681) & (0.0712) &  &  &  &  &  &  &  &  &  &  \\
Complex system & -0.131** & -0.201*** & -0.169*** & -0.209*** & -0.136** &  &  &  &  &  &  &  &  &  &  \\
 & (0.0576) & (0.0578) & (0.0590) & (0.0595) & (0.0633) &  &  &  &  &  &  &  &  &  &  \\
Low dev. complexity & -0.0501 & -0.115* & -0.151** & -0.126* & -0.268*** &  &  &  &  &  &  &  &  &  &  \\
 & (0.0618) & (0.0630) & (0.0662) & (0.0668) & (0.0724) &  &  &  &  &  &  &  &  &  &  \\
High dev. complexity & 0.220*** & 0.342*** & 0.192*** & 0.303*** & 0.178*** &  &  &  &  &  &  &  &  &  &  \\
 & (0.0583) & (0.0585) & (0.0586) & (0.0586) & (0.0617) &  &  &  &  &  &  &  &  &  &  \\
New to the world & 0.0782 & 0.180*** & 0.162*** & 0.150*** & 0.175*** &  &  &  &  &  &  &  &  &  &  \\
 & (0.0504) & (0.0504) & (0.0514) & (0.0515) & (0.0543) &  &  &  &  &  &  &  &  &  &  \\
Radical & 0.341*** & 0.380*** & 0.293*** & 0.336*** & 0.386*** &  &  &  &  &  &  &  &  &  &  \\
 & (0.0708) & (0.0722) & (0.0749) & (0.0768) & (0.0844) &  &  &  &  &  &  &  &  &  &  \\
Major improvement & 0.101 & 0.106 & 0.104 & 0.0894 & 0.192** &  &  &  &  &  &  &  &  &  &  \\
 & (0.0679) & (0.0695) & (0.0717) & (0.0741) & (0.0815) &  &  &  &  &  &  &  &  &  &  \\
Sources & 0.103*** & 0.132*** & 0.0872*** & 0.129*** & 0.106*** &  &  &  &  &  &  &  &  &  &  \\
 & (0.0147) & (0.0147) & (0.0152) & (0.0143) & (0.0147) &  &  &  &  &  &  &  &  &  &  \\
Collaboration & -0.111** & -0.173*** & -0.152*** & -0.148*** & -0.0955* &  &  &  &  &  &  &  &  &  &  \\
 & (0.0507) & (0.0512) & (0.0526) & (0.0532) & (0.0561) &  &  &  &  &  &  &  &  &  &  \\
%Textiles & 0.236 & 0.528** & 0.372 & 0.354 & 0.511** &  &  &  &  &  &  &  &  &  &  \\
% & (0.234) & (0.238) & (0.257) & (0.248) & (0.241) &  &  &  &  &  &  &  &  &  &  \\
%Wood & -0.0601 & -0.235 & -0.421** & -0.405* & 0.0536 &  &  &  &  &  &  &  &  &  &  \\
% & (0.171) & (0.185) & (0.202) & (0.207) & (0.193) &  &  &  &  &  &  &  &  &  &  \\
%Pulp \& paper & 0.185 & -0.0580 & -0.00909 & -0.00785 & 0.241 &  &  &  &  &  &  &  &  &  &  \\
% & (0.184) & (0.199) & (0.200) & (0.201) & (0.203) &  &  &  &  &  &  &  &  &  &  \\
%Chemicals & 0.0418 & 0.200 & 0.127 & 0.194 & 0.227 &  &  &  &  &  &  &  &  &  &  \\
% & (0.135) & (0.135) & (0.139) & (0.137) & (0.139) &  &  &  &  &  &  &  &  &  &  \\
Pharmaceuticals & -0.0542 & 0.862*** & 0.691*** & 0.876*** & 0.998*** &  &  &  &  &  &  &  &  &  &  \\
 & (0.173) & (0.182) & (0.170) & (0.175) & (0.178) &  &  &  &  &  &  &  &  &  &  \\
%Plastics \& rubber & 0.143 & -0.00108 & 0.217* & 0.174 & 0.0366 &  &  &  &  &  &  &  &  &  &  \\
% & (0.114) & (0.117) & (0.120) & (0.122) & (0.128) &  &  &  &  &  &  &  &  &  &  \\
%Basic metals & 0.205 & 0.406*** & 0.429*** & 0.764*** & 0.232 &  &  &  &  &  &  &  &  &  &  \\
% & (0.149) & (0.146) & (0.147) & (0.147) & (0.154) &  &  &  &  &  &  &  &  &  &  \\
%Fabricated metals & 0.250** & 0.0752 & 0.152 & 0.135 & 0.0128 &  &  &  &  &  &  &  &  &  &  \\
% & (0.108) & (0.110) & (0.113) & (0.115) & (0.120) &  &  &  &  &  &  &  &  &  &  \\
Machinery & 0.303*** & 0.214*** & 0.306*** & 0.273*** & 0.0518 &  &  &  &  &  &  &  &  &  &  \\
 & (0.0779) & (0.0786) & (0.0805) & (0.0818) & (0.0847) &  &  &  &  &  &  &  &  &  &  \\
Computers & -0.385*** & -0.320*** & -0.453*** & -0.197 & -0.445*** &  &  &  &  &  &  &  &  &  &  \\
 & (0.120) & (0.121) & (0.127) & (0.127) & (0.140) &  &  &  &  &  &  &  &  &  &  \\
%Electrical apparatus & -0.0401 & 0.00837 & 0.128 & 0.222* & -0.0689 &  &  &  &  &  &  &  &  &  &  \\
% & (0.119) & (0.120) & (0.122) & (0.123) & (0.128) &  &  &  &  &  &  &  &  &  &  \\
%Telecommunication eq. & -0.142 & 0.0292 & -0.0496 & 4.76e-05 & -0.184* &  &  &  &  &  &  &  &  &  &  \\
% & (0.101) & (0.100) & (0.104) & (0.105) & (0.110) &  &  &  &  &  &  &  &  &  &  \\
%Electronic eq. & 0.0275 & 0.105 & 0.145* & 0.138 & -0.194** &  &  &  &  &  &  &  &  &  &  \\
% & (0.0847) & (0.0852) & (0.0872) & (0.0882) & (0.0930) &  &  &  &  &  &  &  &  &  &  \\
%Automotive & 0.0932 & 0.167 & 0.367*** & 0.254* & -0.415*** &  &  &  &  &  &  &  &  &  &  \\
% & (0.133) & (0.132) & (0.134) & (0.137) & (0.156) &  &  &  &  &  &  &  &  &  &  \\
%Recycling & 0.0538 & 0.446* & 0.157 & 0.252 & -0.0234 &  &  &  &  &  &  &  &  &  &  \\
% & (0.280) & (0.270) & (0.278) & (0.289) & (0.300) &  &  &  &  &  &  &  &  &  &  \\
%Software & -0.673*** & -0.203* & -0.535*** & -0.445*** & -0.677*** &  &  &  &  &  &  &  &  &  &  \\
% & (0.110) & (0.104) & (0.122) & (0.118) & (0.132) &  &  &  &  &  &  &  &  &  &  \\
%R\&D & -0.117 & 0.852*** & 0.645*** & 0.776*** & 0.741*** &  &  &  &  &  &  &  &  &  &  \\
% & (0.210) & (0.259) & (0.203) & (0.204) & (0.204) &  &  &  &  &  &  &  &  &  &  \\
Biotechnology & 0.134 & 0.343*** & 0.351*** & 0.386*** & 0.371*** &  &  &  &  &  &  &  &  &  &  \\
 & (0.117) & (0.119) & (0.117) & (0.117) & (0.119) &  &  &  &  &  &  &  &  &  &  \\
Coverage & -2.541*** & -1.720**  &  3.207***  & 0.799***  & 0.120 &  &  &  &  &  &  &  &  &  &  \\
 & (0.621) & (0.827) & (0.484) & (0.208)  & (0.366) &  &  &  &  &  &  &  &  &  &  \\
Loss of rights & 2.035*** &  & 0.612*** & 0.273  & 1.594 &  &  &  &  &  &  &  &  &  &  \\
 & (0.510) &  & (0.225) &  (0.227) & (1.131) &  &  &  &  &  &  &  &  &  &  \\
Duration & 1.464** &  &  & 0.0343   &  &  &  &  &  &  &  &  &  &  &  \\
 & (0.661) &  &  & (0.296) &  &  &  &  &  &  &  &  &  &  &  \\
Enforcement & -0.940** &  & -0.442*  & 0.0380  & 1.902* &  &  &  &  &  &  &  &  &  &  \\
 & (0.391) &  & (0.236)  & (0.203) & (1.082)  &  &  &  &  &  &  &  &  &  &  \\
Membership & 2.431*** & 0.800** & 1.782***  & 0.111 & -0.814* &  &  &  &  &  &  &  &  &  &  \\
 & (0.457) &  (0.338) & (0.263)  & (0.203) & (0.481)  & -0.103 &  &  &  &  &  &  &  &  &  \\
Export (log) &  & -0.0913 & 0.0755 & 0.0387 & (0.191) &  &  &  &  &  &  &  &  &  &  \\
 &  & (0.0819) & (0.231) & (0.0747) &  &  &  &  &  &  &  &  &  &  &  \\
Year & -0.00158 & 0.000580 & -0.0636*** & -0.0139** & 0.00134 &  &  &  &  &  &  &  &  &  &  \\
 & (0.00645) & (0.00372) & (0.00646) & (0.00612) & (0.00793) &  &  &  &  &  &  &  &  &  &  \\
Constant & 0.131 & -1.622 & 121.4*** & 25.50** & -6.792 & 1.128*** & 1.163*** & 1.033*** & 0.928*** & 1.473*** & 1.455*** & 1.259*** & 1.359*** & 1.154*** & 1.244*** \\
 & (12.89) & (7.360) & (12.31) & (11.67) & (15.87) & (0.0352) & (0.0363) & (0.0349) & (0.0357) & (0.0445) & (0.0445) & (0.0455) & (0.0417) & (0.0411) & (0.0421) \\
 &  &  &  &  &  &  &  &  &  &  &  &  &  &  &  \\
Observations & 4,460 & 4,460 & 4,460 & 4,460 & 4,460 & 4,460 & 4,460 & 4,460 & 4,460 & 4,460 & 4,460 & 4,460 & 4,460 & 4,460 & 4,460 \\
\hline
\multicolumn{16}{c}{ Standard errors in parentheses} \\
\multicolumn{16}{c}{ *** p$<$0.01, ** p$<$0.05, * p$<$0.1} \\
\end{tabular}
\caption{Multivariate probit regressions for the propensity of Swedish innovators to patent in five countries. Selected sectors shown.}
\label{tab:mvprobit}
\end{table*}

\begin{table*}[hptb]
\footnotesize
\centering
\begin{tabular}{lccc} \hline
 & (1) & (2) & (3) \\
VARIABLES & Baseline & IPR & IPR DID \\ \hline
 &  &  &  \\
Simple product & 0.0954*** & 0.0967*** & 0.0974*** \\
 & (0.0321) & (0.0323) & (0.0323) \\
Complex system & -0.140*** & -0.142*** & -0.142*** \\
 & (0.0299) & (0.0301) & (0.0301) \\
Low dev. complexity & -0.264*** & -0.265*** & -0.264*** \\
 & (0.0333) & (0.0335) & (0.0335) \\
High dev. complexity & 0.296*** & 0.301*** & 0.302*** \\
 & (0.0279) & (0.0281) & (0.0280) \\
New to the world & 0.279*** & 0.284*** & 0.283*** \\
 & (0.0243) & (0.0244) & (0.0244) \\
Radical & 0.584*** & 0.591*** & 0.591*** \\
 & (0.0393) & (0.0395) & (0.0395) \\
Major improvement & 0.339*** & 0.343*** & 0.343*** \\
 & (0.0382) & (0.0384) & (0.0384) \\
Sources & 0.167*** & 0.168*** & 0.167*** \\
 & (0.00633) & (0.00637) & (0.00635) \\
Collaboration & -0.188*** & -0.190*** & -0.189*** \\
 & (0.0262) & (0.0263) & (0.0263) \\
%Textiles & 0.268** & 0.275** & 0.273** \\
% & (0.112) & (0.112) & (0.112) \\
%Wood & -0.561*** & -0.562*** & -0.563*** \\
% & (0.0961) & (0.0965) & (0.0964) \\
%Pulp \& paper & 0.100 & 0.105 & 0.103 \\
% & (0.0885) & (0.0889) & (0.0888) \\
%Chemicals & 0.276*** & 0.283*** & 0.327*** \\
% & (0.0630) & (0.0635) & (0.0717) \\
%Pharmaceuticals & 1.395*** & 1.426*** & 1.462*** \\
% & (0.0649) & (0.0656) & (0.0731) \\
%Plastics \& rubber & 0.0820 & 0.0841 & 0.0820 \\
% & (0.0574) & (0.0578) & (0.0578) \\
%Basic metals & 0.328*** & 0.336*** & 0.334*** \\
% & (0.0699) & (0.0704) & (0.0703) \\
%Fabricated metals & 0.0791 & 0.0821 & 0.0809 \\
% & (0.0548) & (0.0552) & (0.0552) \\
%Machinery & 0.186*** & 0.191*** & 0.191*** \\
% & (0.0386) & (0.0389) & (0.0389) \\
%Computers & -0.738*** & -0.742*** & -0.743*** \\
% & (0.0677) & (0.0680) & (0.0680) \\
%Electrical apparatus & -0.0512 & -0.0479 & -0.0500 \\
% & (0.0594) & (0.0597) & (0.0597) \\
%Telecommunication eq. & -0.286*** & -0.285*** & -0.287*** \\
% & (0.0516) & (0.0519) & (0.0519) \\
%Electronic eq. & -0.282*** & -0.282*** & -0.282*** \\
% & (0.0434) & (0.0437) & (0.0436) \\
%Automotive & -0.110 & -0.107 & -0.108 \\
% & (0.0685) & (0.0689) & (0.0689) \\
%Recycling & 0.351*** & 0.341*** & 0.341*** \\
% & (0.120) & (0.121) & (0.121) \\
%Software & -1.150*** & -1.155*** & -1.151*** \\
% & (0.0645) & (0.0648) & (0.0647) \\
%R\&D & 0.936*** & 0.943*** & 0.987*** \\
% & (0.0888) & (0.0893) & (0.0975) \\
Coverage &  & 0.0771 & 0.378*** \\
 &  & (0.109) & (0.105) \\
Loss of rights &  & -1.220*** & -1.277*** \\
 &  & (0.102) & (0.102) \\
Duration &  & 1.932*** & 1.931*** \\
 &  & (0.193) & (0.195) \\
Enforcement &  & 0.0530 & 0.113 \\
 &  & (0.0753) & (0.0746) \\
Export (log) &  & 0.404*** & 0.444*** \\
 &  & (0.0357) & (0.0354) \\
PCT &  &  & 0.235*** \\
 &  &  & (0.0509) \\
TRIPS &  &  & 0.578*** \\
 &  &  & (0.132) \\
Budapest &  &  & -0.0830 \\
 &  &  & (0.0615) \\
Biotechnology & 0.337*** & 0.346*** & 0.375*** \\
 & (0.0492) & (0.0496) & (0.0546) \\
Membership &  & 1.175*** &  \\
 &  & (0.109) &  \\
Domestic &  & -1.948*** &  \\
 &  & (0.452) &  \\
Constant & -5.284*** & -4.173*** & -3.808*** \\
 & (0.158) & (0.320) & (0.320) \\
Observations & 245,300 & 243,727 & 243,727 \\
Country FE & YES & YES & YES \\
Year FE & YES & YES & YES \\
Pseudo $R^2$ & 0.305 & 0.315 & 0.315 \\
Log-lik & -35104 & -34499 & -34541 \\ \hline
\multicolumn{4}{c}{ Standard errors in parentheses} \\
\multicolumn{4}{c}{ *** p$<$0.01, ** p$<$0.05, * p$<$0.1} \\
\end{tabular}
\caption{Logistic regressions for the propensity to patent in country $i$}
\label{tab:countrylevel}
\end{table*}

\begin{table*}[hptb]
\footnotesize
\centering
\caption{Logistic regressions (log odds) for prediction of significant innovations from patent quality data, without dummies for technology field.}
\label{tab:predictsuper_without}
\begin{tabular}{lccccc} \hline
 & (1) & (2) & (3) & (4) & (5) \\
VARIABLES & LBIO & LBIO adj. & Blockbuster & New to the world & PCA \\ \hline
 &  &  &  &  &  \\
Citations & 0.0332*** & 0.244*** & 0.0157 & 0.0398*** & 0.0398*** \\
 & (0.00405) & (0.0114) & (0.0112) & (0.0110) & (0.0109) \\
Originality & 0.281*** & 0.147* & -1.511*** & -1.770*** & -1.637*** \\
 & (0.0781) & (0.0851) & (0.158) & (0.165) & (0.167) \\
Radicalness & 0.207*** & 0.265*** & 0.432*** & -0.0327 & 0.0308 \\
 & (0.0700) & (0.0756) & (0.154) & (0.171) & (0.169) \\
Renewal (years) & 0.0479*** &  & -0.0229*** & -0.00694 & -0.00471 \\
 & (0.00332) &  & (0.00705) & (0.00790) & (0.00786) \\
Family size & 0.0307*** & 0.114*** & 0.0530*** & 0.0167** & 0.0167** \\
 & (0.00287) & (0.00372) & (0.00604) & (0.00831) & (0.00826) \\
Filing date & -0.0134*** & -0.0200*** & 0.0133** & 0.0559*** & 0.0569*** \\
 & (0.00193) & (0.00173) & (0.00647) & (0.00789) & (0.00792) \\
Constant & 22.99*** & 37.22*** & -27.58** & -112.8*** & -114.9*** \\
 & (3.863) & (3.446) & (12.95) & (15.82) & (15.87) \\
 &  &  &  &  &  \\
Observations & 74,465 & 25,879 & 6,147 & 6,171 & 6,177 \\
R-squared & 0.0252 & 0.0986 & 0.0300 & 0.0381 & 0.0332 \\
 Tech. field dummies & NO & NO & NO & NO & NO \\ \hline
\multicolumn{6}{c}{ Standard errors in parentheses} \\
\multicolumn{6}{c}{ *** p$<$0.01, ** p$<$0.05, * p$<$0.1} \\
\end{tabular}
\end{table*}

\begin{table*}[hptb]
\footnotesize
\centering
\caption{Logistic regressions (log odds) for prediction of significant innovations from patent quality data, with dummies for technology field.}
\label{tab:predictsuper}
\begin{tabular}{lccccc} \hline
 & (1) & (2) & (3) & (4) & (5) \\
VARIABLES & LBIO & LBIO adj. & Blockbuster & New to the world & PCA \\ \hline
 &  &  &  &  &  \\
Citations & 0.0437*** & 0.254*** & 0.0479*** & 0.0973*** & 0.0851*** \\
 & (0.00430) & (0.0121) & (0.0151) & (0.0154) & (0.0149) \\
Originality & 0.304*** & 0.562*** & -0.660*** & -1.048*** & -0.846*** \\
 & (0.0825) & (0.0941) & (0.234) & (0.251) & (0.253) \\
Radicalness & 0.0832 & -0.157* & 0.249 & 0.118 & 0.240 \\
 & (0.0734) & (0.0831) & (0.207) & (0.231) & (0.228) \\
Renewal (years) & 0.0499*** &  & -0.0164* & 0.0324*** & 0.0392*** \\
 & (0.00343) &  & (0.00953) & (0.0111) & (0.0110) \\
Family size & 0.0415*** & 0.160*** & 0.0627*** & 2.33e-05 & -0.000826 \\
 & (0.00340) & (0.00465) & (0.00852) & (0.0110) & (0.0109) \\
Filing date & -0.00477** & -0.0144*** & 0.00165 & 0.0704*** & 0.0732*** \\
 & (0.00200) & (0.00184) & (0.00748) & (0.00879) & (0.00889) \\
Constant & 6.140 & 26.00*** & -3.739 & -142.5*** & -148.3*** \\
 & (4.015) & (3.673) & (14.98) & (17.63) & (17.83) \\
 &  &  &  &  &  \\
Observations & 74,436 & 25,865 & 5,975 & 6,095 & 6,171 \\
R-squared & 0.0592 & 0.153 & 0.387 & 0.422 & 0.412 \\
 Tech. field dummies & YES & YES & YES & YES & YES \\ \hline
\multicolumn{6}{c}{ Standard errors in parentheses} \\
\multicolumn{6}{c}{ *** p$<$0.01, ** p$<$0.05, * p$<$0.1} \\
\end{tabular}
\end{table*}

\begin{table*}
\footnotesize
\centering
\begin{tabular}{l c c  c}\hline\hline
\multicolumn{1}{c}{\textbf{Variable}} & \textbf{Mean} & \textbf{Std. Dev.} & \textbf{N}\\ \hline
Non-LBIO & 0.761 & 2.231  & 85730\\
LBIO & 1.572 & 4.417  & 4337\\
Blockbuster & 1.817 & 3.641  & 923\\
New to the world & 1.935 & 3.964  & 795\\
PCA & 1.931 & 3.936  & 807\\
\hline\end{tabular}
\caption{Patent citations received within 7 years by EPO patents with a link to LBIO innovations, blockbuster innovations, and patents not linked to LBIO}
\label{tab:citations}
\end{table*}

\begin{table*}
\footnotesize
\begin{tabular}{l*{1}{ccccc}}
\hline
            & Component 1 & Component 2 & Component 3& Component 4 & Component 5\\
\hline
\makecell{Citations \\ (field and time \\ controls)}&       0.488&       0.261&      -0.698&      -0.445&       0.098\\
                    &            &            &            &            &            \\
Originality         &      -0.152&       0.696&       0.195&      -0.205&      -0.642\\
                    &            &            &            &            &            \\
Radicalness         &      -0.307&       0.633&      -0.089&       0.313&       0.632\\
                    &            &            &            &            &            \\
Renewal             &       0.592&       0.145&      -0.046&       0.755&      -0.238\\
                    &            &            &            &            &            \\
Family size         &       0.542&       0.160&       0.682&      -0.305&       0.350\\
                    &            &            &            &            &            \\
Eigenvalue                &            &            &            &            &            \\
                    &       1.572&       1.411&       0.802&       0.679&       0.535\\
                    \hline
\end{tabular}
\caption{Principal component loadings and eigenvalues for Figs. \ref{fig:PCscatter} and \ref{fig:stripplot_pc1}}
\label{tab:pca}
\end{table*}

\begin{table*}
\footnotesize
\centering
\begin{threeparttable}
\begin{tabular}{p{3cm} p{8cm}}
\hline
Feature & Description \\
\hline
\verb|art_comp| & Artefactual complexity  \\
\verb|dev_comp| & Developmental complexity  \\
novelty & Novelty  \\
\verb|year_delta| & Difference between year of commercialization and the year the patent was granted. \\
 & Patent text fields - for each of the three text fields of the patent: title, abstract and description the same type of calculation was applied to generate a quantitative feature:.  \\
\verb|Title_count| &  A count of how many of the keywords can be detected in the title \\
\verb|Abst_count| &   A count of how many of the keywords can be detected in the abstract \\
\verb|desc_count| &  A count of how many of the keywords can be detected in the description \\
\verb|Title_share| & The ratio between how many of the keywords were found in the field to how many keywords there are. \\
\verb|Abst_share| &  Same as above. \\
\verb|desc_share| &  Same as above. \\
\hline
\multicolumn{2}{c}{} \emph{From the second round the below features were added to the textual information}
\\
\hline
\verb|Fulltext_share|  & Fulltext share. By concatenating the three text fields into the complete text, both above mentioned operations were used to generate two new features. \\
\verb|fulltext_count| & Fulltext count. See above \\
\verb|Top_1-10_field| & These features were created by counting the presence of each keyword in the respective text field and keeping the ten highest counts. The \verb|top_1| contains the most commonly occurring keywords and \verb|top_10| the 10th most commonly occurring. \\
\hline
Vectorizing Names\tnote{*} &  \\
\hline
\verb|Inv_count| & Inventor count. Indicates how many of the names with the respective relationship were found among the patent-inventors. \\
\verb|Cont_count|  & Contact count  \\
\verb|Inv_s_share| &  The share of inventors found in the patent. \\
\verb|cont_s_share| & The share of contact persons found in the patent.  \\
\verb|inv_p_share| &  The share of inventors from the patent that were identified. \\
\verb|cont_p_share| & The share of contact persons from the patent that were identified. \\
\hline
\end{tabular}
 \begin{tablenotes}
            \item[*] In the SWINNO database two types of relationships are recorded between innovation and individuals: as inventors/developers or as a contact person. The former indicate that they have been mentioned in an article as having invented or developed the innovation, and the latter that they have been interviewed/cited by a journal. On the patent side multiple names can be recorded as inventor of the patent.
        \end{tablenotes}
        \end{threeparttable}
\caption{Description of features.}
\label{tab:features}
\end{table*}

\begin{table*}
\centering
\footnotesize
\begin{tabular}{l c c c c c c}
\hline
 & Input &  & Output &  \\
 & Innovations & Patents with classification & Innovation-patent pairs & Predicted matches \\
\hline
Round 1 & 99 & 293 & 195,040 & 9,132 \\
Round 2 & 171 & 470 %470 unique patents From the 300 annotations, 204 were resolved successfully: Introducing 199 new unique patents, 72 new unique innovations
& 194,859 & 11,694 \\
Round 3 & 239 & 2,145  %2145 annotated pairings
& 193,169 & 11,487 \\
1970-1984 &  &  & 122,785 & 2,604 \\
\hline
\end{tabular}
\caption{Basic statistics for each round of training the machine-learning model. the input was a number of innovations used as input and a number of patents with manually validation of the link to an innovation (Yes or No). The results of the methodology is a number of identified potential innovation-patent pairs and a number of suggested matches. Rounds 1-3 are based on innovations for the period 1985-2015. The matches for 1970-1984 are based on the ML model for round 3.}
\label{tab:rounds} 
\end{table*}

\begin{table*}
\centering
\footnotesize
\begin{tabular}{l c c c c c c}
\hline
Round & Model & Accuracy & $F_1$-score & FP(Type-I) & FN(Type-II) & N \\
\hline
1 & RandomForest & 0.8077 & 0.7999 & 0.3103 & 0.0435 & 293 \\
2 & RandomForest & 0.7619 & 0.7826 & 0.28 & 0.1765 & 209 \\
3 & RandomForest & 0.7882 & 0.7151 & 0.2237 & 0.2113 & 885 \\
3 & MLP & 0.847 & 0.795 & 0.2623 & 0.0796 & 885 \\
\hline
\end{tabular}
\caption{Statistics for each round and model. Accuracy, F1-score and shares of false positives (FP) and false negatives (FN). Accuracy is calculated as the share of true positives (TP) and true negatives (TN) in the total number of predictions. False positives (negatives) are calculated as shares in the total number of positives (negatives). The $F_1$-score is calculated as $f1 = {2 TP}/\left({2 TP + (FP+FN) }\right)$. 
}
\label{tab:rounds-stats} 
\end{table*}

\clearpage
\newpage
\DeclareFieldFormat{labelnumber}{
  \ifinteger{#1}
    {\number\numexpr#1+37\relax}
    {#1}}
    
\printbibliography[resetnumbers=false]

@article{abrams2018,
author = {D. S. Abrams and U. Akcigit and J. Grennan},
journal = {National Bureau of Economic Research},
title = {Patent value and citations: Creative destruction or strategic disruption?},
year = {2018},
note = {No. w19647},
}

@article{stiglitz2007,
  title={Economic foundations of intellectual property rights},
  author={Stiglitz, Joseph E},
  journal={Duke Law Journal},
  volume={57},
  pages={1693-1724},
  year={2007},
  publisher={HeinOnline}
}

@book{jaffe2011,
  title={Innovation and its discontents: How our broken patent system is endangering innovation and progress, and what to do about it},
  author={Jaffe, Adam B and Lerner, Josh},
  year={2011},
  publisher={Princeton University Press}
}

@incollection{arundel2013,
  title={History of the community innovation survey},
  author={Arundel, Anthony and Smith, Keith},
  booktitle={Handbook of innovation indicators and measurement},
  pages={60--87},
  year={2013},
  publisher={Edward Elgar Publishing}
}

@article{moser2012,
  title={Innovation without patents: Evidence from World’s Fairs},
  author={Moser, Petra},
  journal={The Journal of Law and Economics},
  volume={55},
  number={1},
  pages={43--74},
  year={2012},
  publisher={University of Chicago Press Chicago, IL}
}

@article{moser2005,
  title={How do patent laws influence innovation? Evidence from nineteenth-century world's fairs},
  author={Moser, Petra},
  journal={American economic review},
  volume={95},
  number={4},
  pages={1214--1236},
  year={2005}
}

@book{granstrand2000,
author={Granstrand, Ove},
title={The Economics and Management of Intellectual Property: Towards intellectual capitalism},
publisher={Edward Elgar},
address={Cheltenham},
year={2000},
edition={Paperback ed.},
}

@article{lerner2009,
  title={The empirical impact of intellectual property rights on innovation: Puzzles and clues},
  author={Lerner, Josh},
  journal={American Economic Review},
  volume={99},
  number={2},
  pages={343--348},
  year={2009}
}

@article{lerner2022,
  title={The use and misuse of patent data: Issues for finance and beyond},
  author={Lerner, Josh and Seru, Amit},
  journal={The Review of Financial Studies},
  volume={35},
  number={6},
  pages={2667--2704},
  year={2022},
  publisher={Oxford University Press}
}

@article{holgersson2018,
  title={The evolution of intellectual property strategy in innovation ecosystems: Uncovering complementary and substitute appropriability regimes},
  author={Holgersson, Marcus and Granstrand, Ove and Bogers, Marcel},
  journal={Long Range Planning},
  volume={51},
  number={2},
  pages={303--319},
  year={2018},
  publisher={Elsevier}
}

@book{oecd2018,
   author = "OECD and Eurostat",
   title = "Oslo Manual 2018",
   year = "2018",
   pages = 256,
   url = "https://www.oecd-ilibrary.org/content/publication/9789264304604-en",
   doi = "https://doi.org/https://doi.org/10.1787/9789264304604-en" 
}

@article{arundel1998,
author = {A. Arundel and I. Kabla},
journal = {Research Policy \bf},
number = {2},
pages = {127-141},
title = {What percentage of innovations are patented? Empirical estimates for European firms},
volume = {27},
year = {1998},
}

@incollection{have2009,
  title={Innovation as objective: The SFINNO approach},
  author={van der Have, Robert and Saarinen, Jani and Pesonen, Pekka and Rilla, Nina},
  booktitle={Changes in Innovation},
  pages={9--19},
  year={2009},
  publisher={Springer}
}

@book{soumitra2020,
  title={Global innovation index 2020: who will finance innovation?},
  author={Soumitra, Dutta and Lanvin, Bruno and Wunsch-Vincent, Sacha and others},
  year={2020},
  publisher={WIPO}
}

@article{cohen2000,
author = {W. M. Cohen and R. R. Nelson and J. P. Walsh},
journal = {NBER no. 7552},
title = {Protecting their intellectual assets: appropriability conditions and why US manufacturing firms patent (or not)},
type = {working paper},
year = {2000},
}

@article{criscuolo2008,
author = {P. Criscuolo and B. Verspagen, (},
journal = {Research Policy},
number = {10},
pages = {1892-1908},
title = {Does it matter where patent citations come from? Inventor vs. examiner citations in European patents},
volume = {37},
year = {2008},
}

@article{dernis2001,
author = {H. Dernis and D. Guellec and B. van Pottelsberghe de la Potterie},
journal = {STI Rev.}, 
publisher = {Organisation for Economic Co-operation and Development, Paris, France},
title = {Using patent counts for cross-country comparisons of technology output},
year = {2001},
number ={27},
pages ={129--146},
}

@article{dziallas2019,
author = {M. Dziallas and K. Blind},
journal = {Technovation},
pages = {3-29},
title = {Innovation indicators throughout the innovation process: An extensive literature analysis},
volume = {80},
year = {2019},
}

@article{fontana2013,
author = {R. Fontana and A. Nuvolari and H. Shimizu and A. Vezzulli},
journal = {Research Policy},
number = {10},
pages = {1780-1792},
title = {Reassessing patent propensity: Evidence from a dataset of R \& D awards, 1977–2004},
volume = {42},
year = {2013},
}

@article{gambardella2008,
author = {A. Gambardella and D. Harhoff and B. Verspagen, (},
journal = {European Management Review},
number = {2},
pages = {69-84},
title = {The value of European patents},
volume = {5},
year = {2008},
}

@article{ginarte1997,
  title={Determinants of patent rights: A cross-national study},
  author={Ginarte, Juan C and Park, Walter G},
  journal={Research Policy},
  volume={26},
  number={3},
  pages={283--301},
  year={1997},
  publisher={Elsevier}
}

@article{park2008,
  title={International patent protection: 1960--2005},
  author={Park, Walter G},
  journal={Research Policy},
  volume={37},
  number={4},
  pages={761--766},
  year={2008},
  publisher={Elsevier}
}

@book{wipo2021,
author = {{W}{I}{P}{O}},
title = {Global Innovation Index 2021: Tracking Innovation through the COVID-19 Crisis},
year = {2021},
address ={Geneva},
publisher={{W}orld {I}ntellectual {P}roperty {O}rganization},
}

@article{granstrand2012,
author = {O. Granstrand and M. Holgersson, (},
journal = {International Journal of Intellectual Property Management},
number = {2},
pages = {169-198},
title = {The anatomy of rise and fall of patenting and propensity to patent: the case of Sweden},
volume = {5},
year = {2012},
}

@article{hall2005,
author = {B. H. Hall and A. Jaffe and M. Trajtenberg},
journal = {RAND Journal of Economics},
number = {1},
pages = {16-38},
title = {Market value and patent citations},
volume = {36},
year = {1996},
unidentified = {p},
}

@article{heller1998,
author = {M. A. Heller and R. S. Eisenberg},
journal = {Science},
number = {5364},
pages = {698-701},
title = {Can patents deter innovation? The anticommons in biomedical research},
volume = {280},
year = {1998},
}

@article{higham2021,
author = {K. Higham and G. De Rassenfosse and A. B. Jaffe},
journal = {Research Policy},
number = {4},
pages = {104215},
title = {Patent quality: Towards a systematic framework for analysis and measurement},
volume = {50},
year = {2021},
}

@book{jaffe2002,
address = {Cambridge, Mass.},
author = {A. B. Jaffe and M. Trajtenberg, (},
publisher = {MIT Press},
title = {Patents, Citations, and Innovations: A Window on the Knowledge Economy,},
year = {2002},
}

@article{johansson2022,
author = {M. Johansson and J. Nyqvist and J. Taalbi},
journal = {SSRN},
title = {Linking innovation to patents - a machine learning assisted method},
year = {2022},
url = {https://dx.doi.org/10.2139/ssrn.4127194},
}

@article{kander2019,
author = {A. Kander and J. Taalbi and J. Oksanen and K. Sj{\"o}{\"o} and N. Rilla},
journal = {Scandinavian Economic History Review \bf},
number = {1},
pages = {47-70},
title = {Innovation trends and industrial renewal in Finland and Sweden, 1970-2013},
volume = {67},
year = {2019},
}

@article{kim2004,
author = {J. Kim and G. Marschke},
journal = {Economics of Innovation and New Technology},
number = {6},
pages = {543-558},
title = {Accounting for the recent surge in US patenting: changes in R \& D expenditures, patent yields, and the high tech sector},
volume = {13},
year = {2000},
}

@incollection{Kleinknecht1993,
author = {A. Kleinknecht and J. O. Reijnen and W. Smits, (},
booktitle = {New concepts in innovation output measurement},
pages = {42-84},
title = {Collecting literature based innovation output indicators. The experience in the Netherlands'},
year = {1993},
}

@article{edquist2018,
  title={On the meaning of innovation performance: Is the synthetic indicator of the Innovation Union Scoreboard flawed?},
  author={Edquist, Charles and Zabala-Iturriagagoitia, Jon Mikel and Barbero, Javier and Zof{\'\i}o, Jose Luis},
  journal={Research Evaluation},
  volume={27},
  number={3},
  pages={196--211},
  year={2018},
  publisher={Oxford University Press}
}

@article{roach2013,
author = {M. Roach and W. M. Cohen, (},
journal = {Management Science},
number = {2},
pages = {504-525},
title = {Patent citations as a measure of knowledge flows from public research. Lens or prism?},
volume = {59},
year = {2013},
}

@book{squicciarini2013,
author = {M. Squicciarini and H. Dernis and C. Criscuolo},
publisher = {OECD Publishing},
title = {Measuring Patent Quality. Indicators of Technological and Economic Value},
year = {2013},
}

@article{taalbi2017what,
author = {J. Taalbi},
journal = {Research Policy},
number = {8},
pages = {1437-1453},
title = {What drives innovation? Evidence from economic history},
volume = {48},
year = {2017},
}

@article{trajtenberg1990,
author = {M. Trajtenberg},
journal = {The Rand Journal of Economics},
number = {1},
pages = {172-187},
title = {A penny for your quotes: patent citations and the value of innovations},
volume = {21},
year = {1990},
unidentified = {p},
}

@book{acs1990SI,
author = {Z. J. Acs and D. B. Audretsch},
publisher = {MIT Press},
title = {Innovation and Small Firms},
year = {1990},
address = {Cambridge, Mass.},
}

@article{blind2009SI,
  title={The influence of strategic patenting on companies’ patent portfolios},
  author={Blind, Knut and Cremers, Katrin and Mueller, Elisabeth},
  journal={Research Policy},
  volume={38},
  number={2},
  pages={428--436},
  year={2009},
  publisher={Elsevier},
}

@article{scikit2011SI,
title={Scikit-learn: Machine Learning in {P}ython},
author = {Pedregosa, F. and Varoquaux, G. and  Gramfort, A. and Michel, V. and Thirion, B. and Grisel, O. and Blondel, M. and Prettenhofer, P. and Weiss, R. and Dubourg, V. and Vanderplas, J. and Passos, A. and Cournapeua, D. and Brucher, M. and Perrot, M. and Duchesnay, E. }, 
journal = {Journal of Machine Learning Research}, 
year ={2011},
volume ={12},
pages ={2825--2830},
}

@article{alegrevidal2004SI,
author = {J. Alegre-Vidal and R. Lapiedra-Alcam\'i and R. Chiva-G\'omez},
journal = {Research Policy},
number = {5},
pages = {829-839},
title = {Linking operations strategy and product innovation: an empirical study of Spanish ceramic tile producers},
volume = {33},
year = {2004},

}

@article{arundel1998SI,
author = {A. Arundel and I. Kabla},
journal = {Empirical estimates for European firms, Research Policy \bf},
number = {2},
pages = {127-141},
title = {What percentage of innovations are patented?},
volume = {27},
year = {1998},

}

@article{breschi2000SI,
author = {S. Breschi and F. Malerba and L. Orsenigo},
journal = {The Economic Journal},
number = {463},
pages = {388-410},
title = {Technological regimes and Schumpeterian patterns of innovation},
volume = {110},
year = {2000},

}

@article{brouwer1999SI,
  title={Innovative output, and a firm's propensity to patent: An exploration of CIS micro data},
  author={Brouwer, Erik and Kleinknecht, Alfred},
  journal={Research Policy},
  volume={28},
  number={6},
  pages={615--624},
  year={1999},
  publisher={Elsevier},

}

@techreport{cohen2000SI,
author = {W. M. Cohen and R. R. Nelson and J. P. Walsh},
institution = {NBER},
title = {Protecting their intellectual assets: appropriability conditions and why US manufacturing firms patent (or not)},
type = {working paper},
year = {2000},
unidentified = {(no. 7552)},

}

@article{coombs1996SI,
author = {R. Coombs and P. Narandren and A. Richards},
journal = {Research Policy},
number = {3},
pages = {403-413},
title = {A literature-based innovation output indicator},
volume = {25},
year = {1996},

}

@book{edwards1984SI,
address = {Washington, D.C},
author = {K. L. Edwards and T. J. Gordon},
publisher = {The Futures Group and U.S. Small Business Administration},
title = {Characterization of innovations introduced on the {U}.S. market in 1982},
year = {1984},

}

@article{fontana2013SI,
author = {R. Fontana and A. Nuvolari and H. Shimizu and A. Vezzulli},
journal = {Research Policy},
number = {10},
pages = {1780-1792},
title = {`Reassessing patent propensity: Evidence from a dataset of R \& D awards, 1977–2004'},
volume = {42},
year = {2013},

}

@article{frank1997SI,
author = {S. A. Frank},
journal = {Evolution},
number = {6},
pages = {1712-1729},
title = {The Price equation, Fisher's fundamental theorem, kin selection, and causal analysis},
volume = {51},
year = {1997},
unidentified = {p},

}

@article{ginarte1997SI,
  title={Determinants of patent rights: A cross-national study},
  author={Ginarte, Juan C and Park, Walter G},
  journal={Research Policy},
  volume={26},
  number={3},
  pages={283--301},
  year={1997},
  publisher={Elsevier},

}

@article{park2008SI,
  title={International patent protection: 1960--2005},
  author={Park, Walter G},
  journal={Research Policy},
  volume={37},
  number={4},
  pages={761--766},
  year={2008},
  publisher={Elsevier},
  
}

@article{Grawe2009SI,
author = {S. J. Grawe},
journal = {The International Journal of Logistics Management},
number = {3},
pages = {360-377},
title = {Logistics innovation: a literature-based conceptual framework},
volume = {20},
year = {2009},

}

@article{greve2003aSI,
author = {H. Greve},
journal = {Academy of Management Journal},
number = {6},
pages = {685-702},
title = {A behavioral theory of RD expenditures and innovations: Evidence from shipbuilding},
volume = {46},
year = {2003},

}

@article{derassenfosse2021,
  title={Decentralising the patent system},
  author={De Rassenfosse, Gaetan and Higham, Kyle},
  journal={Government Information Quarterly},
  volume={38},
  number={2},
  pages={101559},
  year={2021},
  publisher={Elsevier}
}

@article{cohen2016,
  title={The growing problem of patent trolling},
  author={Cohen, Lauren and Gurun, Umit G and Kominers, Scott Duke},
  journal={Science},
  volume={352},
  number={6285},
  pages={521--522},
  year={2016},
  publisher={American Association for the Advancement of Science}
}

@article{heller1998SI,
author = {M. A. Heller and R. S. Eisenberg},
journal = {Science},
number = {5364},
pages = {698-701},
title = {Can patents deter innovation? The anticommons in biomedical research},
volume = {280},
year = {1998},

}

@article{johansson2022SI,
author = {M. Johansson and J. Nyqvist and J. Taalbi},
journal = {SSRN},
title = {Linking innovation to patents - a machine learning assisted method},
year = {2022},
url = {https://dx.doi.org/10.2139/ssrn.4127194},

}

@article{kander2019SI,
author = {A. Kander and J. Taalbi and J. Oksanen and K. Sj{\"o}{\"o} and N. Rilla},
journal = {Scandinavian Economic History Review \bf},
number = {1},
pages = {47-70},
title = {Innovation trends and industrial renewal in Finland and Sweden, 1970-2013},
volume = {67},
year = {2019},

}

@article{kleinknecht2002SI,
author = {A. Kleinknecht and K. van Montfort and E. Brouwer, (},
journal = {Economics of Innovation and New Technology},
number = {2},
pages = {109-121},
title = {The non-trivial choice between innovation indicators},
volume = {11},
year = {2002},

}

@article{makkonen2013SI,
author = {T. Makkonen and R. P. van der Have},
journal = {Scientometrics},
number = {1},
pages = {247-262},
title = {Benchmarking regional innovative performance: composite measures and direct innovation counts},
volume = {94},
year = {2013},
unidentified = {p},

}

@techreport{palmberg1999SI,
author = {C. Palmberg and A. Lepp\"alahti and T. Lemola and H. Toivanen},
institution = {Espoo: VTT},
title = {Towards a better understanding of innovation and industrial renewal in Finland: A new perspective},
year = {1999},

}

@phdthesis{Saarinen2005SI,
author = {J. Saarinen},
school = {Diss. Lund Univ},
title = {Innovations and Industrial Performance in Finland, 1945-1998},
year = {2005},

}

@article{santarelli1996SI,
author = {E. Santarelli and R. Piergiovanni},
journal = {Research Policy},
number = {5},
pages = {689-711},
title = {Analyzing literature-based innovation output indicators: the Italian experience},
volume = {25},
year = {1996},

}

@book{sedig2002SI,
author = {K. Sedig and D. M. Olson},
publisher = {Swedish Institute [Svenska institutet]},
title = {Swedish innovations},
year = {2002},

}

@article{shannon1948SI,
author = {C. E. Shannon},
journal = {The Bell system technical journal},
pages = {379-423},
title = {A mathematical theory of communication},
volume = {27},
number = {3},
year = {1948},

}

@article{Sjoo2013SI,
author = {K. Sj{\"o}{\"o} and J. Taalbi and A. Kander and J. Ljungberg},
journal = {Lund Papers in Economic History},
pages = {1970-2007},
title = {SWINNO - a database of Swedish innovations},
volume = {133},
year = {2014},

}

@book{squicciarini2013SI,
author = {M. Squicciarini and H. Dernis and C. Criscuolo},
publisher = {OECD Publishing},
title = {Measuring Patent Quality. Indicators of Technological and Economic Value},
year = {2013},

}

@article{lanjouw2004SI,
  title={Patent quality and research productivity: Measuring innovation with multiple indicators},
  author={Lanjouw, Jean O and Schankerman, Mark},
  journal={The economic journal},
  volume={114},
  number={495},
  pages={441--465},
  year={2004},
  publisher={Oxford University Press Oxford, UK}
}

@article{taalbi2017whatSI,
author = {J. Taalbi},
journal = {Research Policy},
number = {8},
pages = {1437-1453},
title = {What drives innovation? Evidence from economic history},
volume = {48},
year = {2017},

}

@article{taalbi2021SI,
author = {J. Taalbi},
journal = {Environmental Innovation and Societal Transitions},
number = {7},
pages = {222-248},
title = {Innovation in the long run. Perspectives on technological transitions in Sweden, 1908-2016.},
volume = {40},
year = {2021},

}

@article{walker2002SI,
author = {R. M. Walker and E. Jeanes and R. Rowlands},
journal = {Public Administration},
number = {1},
pages = {201-214},
title = {Measuring Innovation-Applying the Literature-Based Innovation Output Indicator to Public Services},
volume = {80},
year = {2002},

}

@article{wallmark1991SI,
author = {J. T. Wallmark and D. H. McQueen},
journal = {Research Policy},
number = {4},
pages = {325-344},
title = {One hundred major Swedish technical innovations, from 1945 to 1980},
volume = {20},
year = {1991},

}
\end{refsection}

\end{document}